\documentclass[sn-vancouver,Numbered]{sn-jnl}

\usepackage{graphicx}%
\usepackage{multirow}%
\usepackage{amsmath,amssymb,amsfonts}%
\usepackage{amsthm}%
\usepackage{mathrsfs}%
\usepackage[title]{appendix}%
\graphicspath{{./Figures/}}
\usepackage{subcaption}
\usepackage{xcolor}%
\usepackage{tabularx}
\usepackage{tabularray}
\usepackage{qcircuit}
\usepackage{textcomp}%
\usepackage{manyfoot}%
\usepackage{braket}
\usepackage{colortbl}
\usepackage{dcolumn}
\usepackage{booktabs}%
\usepackage{algorithm}%
\usepackage{algorithmicx}%
\usepackage{algpseudocode}%
\usepackage{listings}%

\begin{document}

\title[Article Title]{A Quantum Approach to Synthetic Minority Oversampling Technique (SMOTE)}

\author*[1]{\fnm{Nishikanta} \sur{Mohanty}}\email{nishikanta.m.mohanty@student.uts.edu.au}

\author[2]{\fnm{Bikash K.} \sur{Behera}}\email{bikas.riki@gmail.com}

\author[1]{\fnm{Christopher} \sur{Ferrie}}\email{Christopher.Ferrie@uts.edu.au}

\author[2]{\fnm{Pravat} \sur{Dash}}\email{pravat.dash@outlook.com}

\affil[1]{\orgdiv{Centre for Quantum Software and Information}, \orgname{University of Technology Sydney}, \orgaddress{\street{15 Broadway, Ultimo}, \city{Sydney}, \postcode{2007}, \state{NSW}, \country{Australia}}}

\affil[2]{\orgdiv{} \orgname{Bikash's Quantum (OPC) Pvt. Ltd.}, \orgaddress{\street{Balindi}, \city{Mohanpur}, \postcode{741246}, \state{WB}, \country{India}}}

\abstract{The paper proposes the Quantum-SMOTE method, a novel solution that uses quantum computing techniques to solve the prevalent problem of class imbalance in machine learning datasets. Quantum-SMOTE, inspired by the Synthetic Minority Oversampling Technique (SMOTE), generates synthetic data points using quantum processes such as swap tests and quantum rotation. The process varies from the conventional SMOTE algorithm's usage of K-Nearest Neighbors (KNN) and Euclidean distances, enabling synthetic instances to be generated from minority class data points without relying on neighbor proximity. The algorithm asserts greater control over the synthetic data generation process by introducing hyperparameters such as rotation angle, minority percentage, and splitting factor, which allow for customization to specific dataset requirements. Due to the use of a compact swap test, the algorithm can accommodate a large number of features. Furthermore, the approach is tested on a public dataset of TelecomChurn and evaluated alongside two prominent classification algorithms, Random Forest and Logistic Regression, to determine its impact along with varying proportions of synthetic data.}


\maketitle

\section{Introduction}\label{sec1}

\subsection{Unbalanced Classification}
Unbalanced classification is a prevalent problem in machine learning \cite{haixiang_learning_2017, blaszczyk_framework_2021}, especially when the classes in a dataset are not represented evenly. Due to this imbalance, models may be biased towards the dominant class, frequently at the price of adequately forecasting the minority class. Such scenarios are common in real-world applications such as fraud detection in banking, insurance, and retail industries, detecting spam in email content, and predicting customer churn in Telecom, where the class of interest is usually underrepresented. To mitigate the problem of unbalanced classes, multiple techniques are used across industries, out of which Synthetic Minority Oversampling Techniques (SMOTE) \cite{wang_research_2021, chawla_smote_2002} are quite popular.

\subsection{Overview of SMOTE}

SMOTE is a statistical method used to augment the number of instances in a dataset in a balanced manner. The technique was first presented by Chawla et al. \cite{chawla_smote_2002}, whose main objective is to tackle the issue of imbalanced datasets, namely in the realm of classification. Imbalanced datasets are common in many real-world circumstances, where the frequency of instances belonging to a certain class is much lower than the others. The disparity may result in unsatisfactory performance of classification models since they have a tendency to exhibit bias towards the dominant class. SMOTE resolves this problem by generating artificial samples from the underrepresented class.

\subsection{Existing works on SMOTE}

During our study and implementation of the SMOTE technique and its modifications, we have come across academic papers authored by other researchers that explore the progress and real-world uses of this algorithm \cite{fernandez_smote_2018, asi4010018, Seiffert2010RUSBoostAH}. Research on the incorporation of SMOTE into ensemble learning approaches has been a substantial focus. The combination seeks to use the advantages of both techniques in order to enhance the classification performance on datasets with uneven distribution. The use of SMOTEBoost \cite{smoteboost_2003}, and RusBoost highlights the significance of SMOTE in ensemble learning techniques. Moreover, current research is underway in the domain of image classification with a specific emphasis on the use of SMOTE \cite{app13064006}.

\subsection{Purpose and Scope}
Since SMOTE is a widely used technique in machine learning to address unbalanced classification, we believe that a quantum computing approach will greatly enhance its efficiency in quantum machine learning applications. Since quantum computing is greatly useful in problems related to high dimensional datasets, A SMOTE algorithm in quantum machine Learning will be of significant value. In this paper, we propose a novel method of generating synthetic data points by using the quantum swap test and quantum rotations, which can be used to increase the number of minority class representatives in a large dataset and help reduce bias in classification models. We have also applied the method to a publicly available dataset named Telco Customer Churn \cite{telco_churn_dataset_kaggle} used for telecom churn classification and recorded the results. 

\subsection{Organization}
The paper is structured in the following manner. Section \ref{Background} explores the core mathematical principles, including the Basic Concept, several versions of the SMOTE algorithm, and the K-Means Clustering technique. Section \ref{Methodology} presents an examination of the development of SMOTE utilising quantum techniques, namely the use of the swap test and rotation principles. This is followed by analyzing the outcomes obtained by applying these concepts to actual data. Section \ref{Case Study and Results} involves the application of the quantum SMOTE algorithm to a real-world dataset. This process comprises data preparation, clustering, and the production of synthetic data using the SMOTE method. We utilise the SMOTE technique on the telecom data, varying the proportions of the minority class to 30\%, 40\%, and 50\%, respectively. In Section \ref{Inferences}, we provide a summary of the results and model parameters of the classification Models, which elucidate the effects of Quantum SMOTE.

\section{Background} \label{Background}

\subsection{Basic Concept of SMOTE} \label{SMOTE}
SMOTE was proposed way back in 2002 by Chawla et al. \cite{chawla_smote_2002} as a way to address issues with unbalanced classification. The primary objective of the SMOTE algorithm is to generate Synthetic data points from minority classes using K Nearest neighbors and Euclidean distances. The synthetic data points, in turn, increase the population of the minority class in the population, which counters the bias towards the majority class in a classification scenario. SMOTE is widely used and accepted, and since then, multiple variants of SMOTE have been proposed by various researchers. In the below subsections, we will cover the working of the SMOTE algorithm and its Variants. 

\subsection{How SMOTE Works}

SMOTE \cite{chawla_smote_2002} is an over-sampling technique that addresses imbalanced datasets by generating synthetic instances for the minority class instead of just duplicating existing examples. To address the imbalance in class distribution, the minority class is augmented by generating synthetic samples along the line segments connecting the K nearest neighbours of each minority class sample. Neighbours are randomly selected from the K nearest neighbours, based on the desired level of over-sampling. The initial approach used a set of five closest neighbours. For example, when the required over-sampling quantity is 300\%, only three neighbours are selected from the five nearest neighbours, and one sample is created in the direction of each selected neighbour. 

Synthetic samples are produced as follows:
\begin{itemize}
\item[1.] Find the feature vector's closest neighbor and compute the difference between the two.
\item[2.] Pick a uniformly random number between 0 and 1 and multiply it by this difference. 
\item[3.]\ Add the resulting number to the original feature vector. 
\end{itemize}

The result is the random creation of a synthetic point along the line segment between two feature vectors. This method broadens the minority group's density and resolves the decision boundary.

\begin{algorithm}[H]
\caption{SMOTE$(N, A, m)$}
\begin{algorithmic}[1]
\State \textbf{Input:}
\State $N$ = number of samples in the minority class.
\State $A$ = the percentage of SMOTE to be applied.
\State $m$ = number of nearest neighbours to be considered.
\State \textbf{Output:}
\State Generate $(N/100) \times A$ artificial samples for the minority class.

\Procedure{SMOTE}{$N, A, m$}
    \If{Proportion of class $A < 100\%$}
        \State Randomly choose a percentage of the minority class samples to be SMOTEd.
    \EndIf
    \If{$A < 100$}
        \State $N \gets (A/100) \times N$
        \State $A \gets 100$
    \EndIf
    \State $A \gets \text{int}(A/100)$
    \State $numattrs \gets$ total count of attributes
    \State $Sample[][] \gets$ array containing the original minority class samples
    \State $newindex \gets 0$
    \State $Synthetic[][] \gets$ array for creating artificial samples
    \For{$i = 1$ to $N$}
        \State Compute $m$ closest neighbours for $i$ and save indices in $nnarray$
        \State Fill array $A$ with values from $nnarray$ starting at index $i$
    \EndFor
    \State \Call{Populate}{$A, i, nnarray$}
    \While{$A \neq 0$}
        \State Select a random integer from $1$ to $m$ as $nn$
        \For{$attr = 1$ to $numattrs$}
            \State $diff \gets Sample[nnarray[nn]][attr] - Sample[i][attr]$
            \State $gap \gets$ random number between $0$ and $1$
            \State $Synthetic[newindex][attr] \gets Sample[i][attr] + gap \times diff$
        \EndFor
        \State $newindex \gets newindex + 1$
        \State $A \gets A - 1$
    \EndWhile
    \State \Return ``End of Populate''
\EndProcedure
\end{algorithmic}
\end{algorithm}

\subsection{Variants of SMOTE}
As the SMOTE algorithm became popular, multiple variations have been proposed. For the sake of reference, we mention some of them in this section.

\textbf{Borderline-SMOTE}:

Borderline-SMOTE specifically targets the minority class samples that are in close proximity to the boundary with the majority class. The objective is to produce artificial samples at close proximity to the boundary rather than over the whole of the distribution of the minority class \cite{han_borderline-smote_2005}.

\textbf{ADASYN (Adaptive Synthetic Sampling)}:

ADASYN specifically aims to generate synthetic samples for the minority class. However, unlike SMOTE, ADASYN adjusts its approach based on the unique properties of the dataset. It produces additional synthetic data for minority class samples that are more challenging to learn (i.e., those that are incorrectly categorized using the K-Nearest Neighbor method) in contrast to those that are less difficult. The number of artificial samples to be generated for each underrepresented sample is contingent upon the complexity of learning that particular sample \cite{AMASYN_He_2008}.

\textbf{SMOTE-ENN (SMOTE with Edited Nearest Neighbors)}:

SMOTE-ENN \cite{Smote_enn_batista_2004} is a hybrid technique that integrates the concepts of over-sampling and under-sampling to tackle the problem of class imbalance in machine learning. The SMOTE method is used to oversample the minority class, whereas the ENN rule is used for undersampling. SMOTE algorithm creates new samples in the minority class by selecting the K-nearest neighbors from the same class and creating interpolations between the original sample and its neighbors.

Each instance in the dataset undergoes testing by comparing it with its three closest neighbors. If the majority of the neighbors do not have the same class as the instance, the instance is removed. This mostly pertains to the dominant class within skewed datasets. \\
The implementation of SMOTE-ENN involves the following steps: \\
\textit{Initial Step}: Utilise SMOTE technique to oversample the minority class and generate synthetic instances, hence achieving a balanced distribution of classes. \\
\textit{Next}, implement the ENN rule on the dataset that has been oversampled. ENN will exclude instances from both the majority and minority classes that are deemed to be noisy or are located on the boundary between the two classes.\\
\textit{Result}: This integrated method not only resolves the disparity by augmenting the number of instances in the underrepresented category but also enhances the dataset's quality, resulting in a more distinct and less susceptible decision border between the classes, reducing overfitting. This helps in cleaning the space between the majority and minority classes.

\textbf{SMOTE-Tomek Links}:

SMOTE-Tomek Links is a hybrid method that combines the Synthetic Minority Over-sampling Technique (SMOTE) with Tomek Links, an under-sampling technique. This combination is used to mitigate class imbalance in machine learning datasets. 
A pair of examples belonging to contrasting classes are classified as a Tomek Link if they are the closest neighbours of each other. Essentially, they are closely related points, but belong to separate classes.
The objective is to eliminate these Tomek Links in order to enhance the clarity of the class boundaries. Usually, the instance belonging to the majority class in each pair of Tomek Links is eliminated, which helps in minimising the overlap between classes \cite{chawla_smote_2002, I_TOMEK_1976}.

\textbf{SVMSMOTE}:

SVMSMOTE (Support Vector Machine Synthetic Minority Over-sampling Technique) \cite{SVM_SMOTE_Demidova_2017} integrates ideas from Support Vector Machines (SVMs) into SMOTE. SVMSMOTE uses SVMs to detect support vectors among the samples of the minority class. Support vectors are often defined as the data points that are in close proximity to the decision border separating different classes. Within the framework of class imbalance, these minority class samples are often the most crucial ones to prioritise for over-sampling. SVMSMOTE creates synthetic samples in the proximity of the detected support vectors rather than distributing them randomly throughout the whole space of the minority class. The objective of this strategy is to enhance the decision border region where the classifier is prone to uncertainty.

\subsection{K-Means Clustering}
K-means clustering \cite{Lloyd_quantization_1982} is a widely used unsupervised machine learning approach that divides a dataset into K separate and non-overlapping groups. The main goal of the K-means algorithm is to categorise data points into clusters, where each point is assigned to the cluster with the closest average value, which acts as the centre or centroid of the cluster. The technique sequentially allocates data points to the centroid that is closest to them and updates the locations of the centroids by calculating the mean of the points in each cluster. This procedure iterates until convergence, which is achieved when the locations of the centroids no longer exhibit substantial changes or when a predetermined number of iterations is reached. The k-means algorithm is very susceptible to the starting position of centroids, which might result in convergence to local optima. Therefore, it is crucial to do numerous iterations of the algorithm with various initialisations to ensure accurate results. Although K-means is computationally fast and easy to implement, its main strengths lie in its ability to uncover patterns in data, cluster comparable observations, and assist in exploratory data analysis in many domains, such as picture segmentation, customer segmentation, and pattern identification.

\subsection{ROC Curve}
The Receiver Operating Characteristic (ROC) is a commonly used graphical plot to assess the effectiveness of a binary classifier system while the discrimination threshold is adjusted. It is especially advantageous in scenarios where there is a requirement to strike a balance between a true positive rate and a false positive rate.

The True Positive Rate (TPR), often referred to as Sensitivity, Recall, or Probability of Detection, is determined by the formula $TPR = TP / (TP + FN)$, where $TP$ represents the count of true positives and $FN$ represents the count of false negatives.
The False Positive Rate (FPR), often referred to as the Probability of False Alarm, is determined by the formula $FPR = FP / (FP + TN)$, where $FP$ represents the count of false positives and $TN$ represents the count of true negatives.

An ROC curve illustrates the relationship between the true positive rate (TPR) and the false positive rate (FPR) across different threshold values. The $x$-axis corresponds to the False Positive Rate, while the $y$-axis corresponds to the True Positive Rate.

The AUC, or Area Under the Curve, is a metric that quantifies a classifier's capacity to differentiate between classes. It serves as a concise representation of the ROC curve. A model with a higher AUC value indicates superior performance.

\section{Emulating SMOTE Using Quantum} \label{Methodology}
Upon examining the SMOTE algorithm and its modifications as presented by \cite{chawla_smote_2002}, we have adopted a distinct method for oversampling the minority class by using quantum approaches. It is often seen in real-world datasets that the minority class is unevenly distributed in the population. Therefore, producing synthetic data uniformly throughout all distribution zones may not effectively address the issue of bias. Our method entails dynamically segmenting the whole population using clustering methods and thereafter creating synthetic data inside each cluster to achieve the desired minority proportion. The target minority percentage is the overall percentage of minorities in the population following the introduction of synthetic data.

Synthetic data creation requires using quantum rotation to manipulate individual data points from the minority class. This is done by representing each data point as a multidimensional vector and rotating it along the X (or Y or Z)  direction. In the next part, we will get into the specifics of selecting X rotations. The rotation angle is computed as the angle formed between the vector of the minority data point and the centroid vector of the cluster it belongs to. It is important to mention that while determining the angle slice, a relatively tiny angle is used to reduce sudden departures from the initial minority class data point. If there are many synthetic data points to be created, the remaining synthetic data points are obtained by incrementing the angle from the starting value.

The objective of this strategy is to ensure that the created synthetic data points remain within the statistical bounds of their respective cluster while also boosting the density of the minority class. In the following sections, we will provide a comprehensive analysis of the algorithm, rotation, and data creation process.

\begin{figure*}[!ht]
\centering
\begin{subfigure}{0.5\linewidth}
\includegraphics[width=\linewidth]{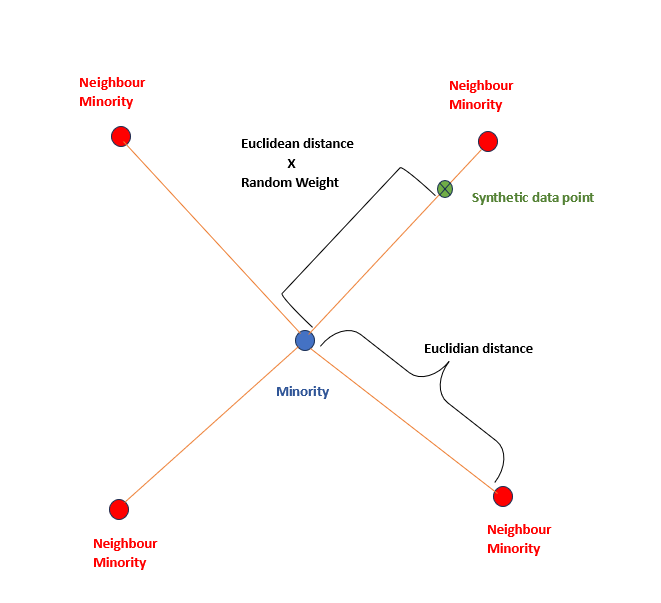} 
\caption{}
\label{classical_SMOTE}
\end{subfigure}\hfill
\begin{subfigure}{0.5\linewidth}
\includegraphics[width=\linewidth]{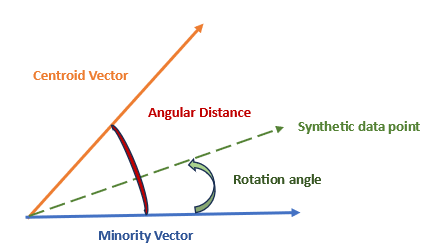} 
\caption{}
\label{Quantum_SMOTE}
\end{subfigure}\hfill
\caption{Plot illustrating different SMOTE mechanisms. (a) Classical SMOTE, (b) Quantum SMOTE.}
\label{SMOTE_Mechanisms}
\end{figure*}

The figures \Ref{SMOTE_Mechanisms} illustrate fundamental difference in Classical and Quantum SMOTE procedures 

\subsection{Swap Test} \label{Swap Test}

The quantum swap test is a quantum procedure used to ascertain the degree of similarity between two quantum states, $\psi$ and $\phi$. The test result quantifies the degree of overlap between the two states, which is directly linked to their inner product $\braket{\psi|\phi}$. Usually, we tackle the swap test in the following manner.

\textit{Setup}: Commence by using a control qubit, normally in the state $\ket{0}$, together with two quantum registers that are in the respective states $\psi$ and $\phi$.

\textit{Hadamard Transformation}: Perform a Hadamard gate operation on the control qubit. This results in the creation of a superposition state, where the control qubit is in a state that is proportional to the sum of $\ket{0}$ and $\ket{1}$.

\textit{Controlled Swap}: Execute a regulated exchange (or Fredkin gate) using the control qubit. When the control qubit is in the state $\ket{1}$, it performs a swap operation on the two quantum registers. Alternatively, it does not alter them.

\textit{Second Hadamard}: Apply a second Hadamard gate to the control qubit.

\textit{Measurement}: Conduct a measurement on the control qubit. If the two quantum states $\ket{\psi}$ and $\ket{\phi}$ are indistinguishable, the control qubit will consistently be seen in the state $\ket{0}$. The likelihood of seeing the state $\ket{0}$ diminishes as the states grow more different.

\textit{Outcome}: The chance of seeing the control qubit in the state $\ket{0}$ after the swap test provides information on the similarity of the two quantum states. More precisely, the likelihood is proportional to the square of the magnitude of their inner product.
The mathematical expression for the probability $P(0)$ of measuring the state $\ket{0}$ is,

\begin{eqnarray}
P_0 = \frac{1}{2}(1 + |\braket{\psi|\phi}|^2).
\end{eqnarray}
From this above expression, $\braket{\psi|\phi}$ can be determined as,

\begin{eqnarray}
\braket{\psi|\phi}=\sqrt{2P_0-1}
\end{eqnarray}
Fig. \ref{Swap test circuit Orig} circuit illustrates the basic swap test.
\begin{figure}
\centering
\Qcircuit @C=3em @R= 3em {
&& \lstick{\ket{0}} & \gate{H} & \ctrl{2} & \gate{H} & \meter \\
&& \lstick{\ket{\psi}} & \qw & \qswap & \qw & \qw  \\
&& \lstick{\ket{\phi}} & \qw & \qswap & \qw & \qw  \\
}
\caption{Swap test circuit.}
\label{Swap test circuit Orig}
\end{figure}

The swap test probability can be defined as,
\begin{eqnarray}
\text{swap\_test\_probability} = 1 - 2 p_0 + p_1
\end{eqnarray}
where $p_0$ and $p_1$  are probabilities of the states $\ket{0}$ and $\ket{1}$ respectively.

\subsubsection{Compact Swaptest} \label{Compact Swaptest}
For the purpose of this paper, we have adopted a modified version of the swap test to find the inner product of our two vectors, namely the centroid and an arbitrary minority data point within the cluster. The procedure is already discussed in the articles \cite{qiskit_medium, Mart_Dissimilarity_2023}. Though the article describes the procedure as a dissimilarity measure and uses it to calculate Euclidian distance, we have used it to calculate the inner product of quantum states and thereby the angular distance. The advantage of this procedure is that it requires less number of qubits $$n=\log _2(M)+1$$ where n is the number of qubits and M is the classical data encoded by amplitude embedding. The procedure is as follows,

We amplitude encode two vectors C (Centroid) and M (Minority) by 

\begin{eqnarray}
C \longrightarrow|C\rangle & =&\frac{1}{|C|} \sum_i C_i\left|q_i\right\rangle \\
M \longrightarrow|M\rangle & =&\frac{1}{|M|} \sum_i M_i\left|q_i\right\rangle
\end{eqnarray}

We define the quantum states $|\psi\rangle$ and $|\phi\rangle$ as:

\begin{eqnarray}
|\psi\rangle&=&\frac{|0\rangle \otimes|C\rangle+|1\rangle \otimes|M\rangle}{\sqrt{2}}  \nonumber \\
|\phi\rangle&=&\frac{|C||0\rangle-|M||1\rangle}{\sqrt{Z}} \nonumber \\
Z&=&|C|^2 + |M|^2
\end{eqnarray}

lets divulge into the details of this circuit

\begin{equation}
\begin{aligned}
& |0\rangle|\phi\rangle|\psi\rangle \\
= & |+\rangle \left(\frac{(C|0\rangle-M|1\rangle)}{\sqrt{Z}}\right)\left(\frac{|0\rangle|C\rangle+|1\rangle|M\rangle}{\sqrt{2}}\right) \\
= & \left(\frac{|0\rangle+|1\rangle}{\sqrt{2}}\right)\left(\frac{C|0\rangle-M|1\rangle}{\sqrt{Z}}\right)\left(\frac{|0\rangle|C\rangle+|1\rangle|M\rangle}{\sqrt{2}}\right) \\
= & \frac{1}{2 \sqrt{Z}}[|0\rangle(C|0\rangle-M|1\rangle)(|0\rangle|C\rangle+|1\rangle|M\rangle) \\
& +|1\rangle(C|0\rangle-M|1\rangle)(|0\rangle|C\rangle+|1\rangle|M\rangle)] \\
= & \frac{1}{2 \sqrt{Z}}[|0\rangle(C|0\rangle|0\rangle|C\rangle+C|0\rangle|1\rangle|M\rangle-M|1\rangle|0\rangle|C\rangle-M|1\rangle|1\rangle|M\rangle) \\
& +|1\rangle(C|0\rangle|0\rangle|C\rangle+C|0\rangle|1\rangle|M\rangle-M|1\rangle|0\rangle|C\rangle-M|1\rangle|1\rangle|M\rangle)]
\end{aligned}
\end{equation}

Applying controlled swap operation

\begin{eqnarray}
&=&\frac{1}{2 \sqrt{Z}} {[|0\rangle(C|0\rangle|0\rangle|C\rangle+C|0\rangle| 1\rangle|M\rangle-M|1\rangle|0\rangle|C\rangle-M|1\rangle|1\rangle|M\rangle) } \nonumber\\
& +&|1\rangle(C|0\rangle|0\rangle|C\rangle+C|1\rangle|0\rangle|M\rangle-M|0\rangle|1\rangle|C\rangle-M|1\rangle|1\rangle|M\rangle]
\end{eqnarray}

Applying Hadamard 
\begin{eqnarray}
&=&\frac{1}{2 \sqrt{Z}} {[|+\rangle(C|0\rangle|0\rangle|C\rangle+C|0\rangle| 1\rangle|M\rangle-M|1\rangle|0\rangle|C\rangle-M|1\rangle|1\rangle|M\rangle) } \nonumber\\
& +&|-\rangle(C|0\rangle|0\rangle|C\rangle+C|1\rangle|0\rangle|M\rangle-M|0\rangle|1\rangle|C\rangle-M|1\rangle|1\rangle|M\rangle]\nonumber\\
&=&\frac{1}{2 \sqrt{2Z}} {[(|0\rangle+|1\rangle)(C|0\rangle|0\rangle|C\rangle+C|0\rangle|1\rangle|M\rangle-M|1\rangle|0\rangle|C\rangle-M|1\rangle|1\rangle|M\rangle) } \nonumber\\
&+&(|0\rangle-|1\rangle)(C|0\rangle|0\rangle|C\rangle+C|1\rangle|0\rangle|M\rangle-M|0\rangle|1\rangle|C\rangle-M|1\rangle|1\rangle|M\rangle] \nonumber \\
&=&\frac{1}{2 \sqrt{2Z}} {[|0\rangle(2 C|0\rangle|0\rangle|C\rangle+(C|0\rangle|1\rangle|M\rangle-M|0\rangle|1\rangle|C\rangle)+(C|1\rangle|0\rangle|M\rangle} \nonumber\\
&-&M|1\rangle|0\rangle|C\rangle) -2 M|1\rangle|1\rangle|M\rangle) \nonumber\\
&+&|1\rangle(C|0\rangle|1\rangle|M\rangle+M|0\rangle|1\rangle|C\rangle-M|1\rangle|0\rangle|C\rangle-C|1\rangle|0\rangle|M\rangle]
\end{eqnarray}

The probability of 0 can be calculated as,

\begin{eqnarray} \label{inner product of vectors}
P_0&=&\frac{1}{8 Z}|(2 C|0\rangle|0\rangle|C\rangle+(C|0\rangle|1\rangle|M\rangle-M|0\rangle|1\rangle|C\rangle)\nonumber\\
&+&(C|1\rangle|0\rangle|M\rangle -M|1\rangle|0\rangle|C\rangle)-2 M|1\rangle|1\rangle|M\rangle)|^2\nonumber\\
&=&\frac{1}{8 Z}|(2 C|0\rangle|0\rangle|C\rangle+|0\rangle|1\rangle(C|M\rangle-M|C\rangle)\nonumber\\
&+&|1\rangle|0\rangle(C|M\rangle -M|C\rangle)-2 M|1\rangle|1\rangle|M\rangle)|^2\nonumber\\
&=&\frac{1}{8 Z}|(2 C|0\rangle|0\rangle|C\rangle+(|0\rangle|1\rangle+|1\rangle|0\rangle)(C|M\rangle-M|C\rangle)-2 M|1\rangle|1\rangle|M\rangle)|^2\nonumber\\
&=&\frac{1}{8 Z}|(2 C|0\rangle|0\rangle|C\rangle-2 M|1\rangle|1\rangle|M\rangle)+(|0\rangle|1\rangle+|1\rangle|0\rangle)(C|M\rangle-M|C\rangle)|^2\nonumber\\
&=&\frac{1}{8 Z}(|2 C|0\rangle|0\rangle|C\rangle-2 M|1\rangle|1\rangle|M\rangle|^2+||0\rangle|1\rangle+|1\rangle|0\rangle|^2|C|M\rangle-M|C\rangle|^2)\nonumber\\
&=&\frac{1}{8 Z}(4C^2+4M^2+||0\rangle|1\rangle+|1\rangle|0\rangle|^2|C|M\rangle-M|C\rangle|^2)\nonumber\\
&=&\frac{1}{8 Z}(4Z+2(C^2+M^2-2CM\braket{M|C}))\nonumber\\
&=&\frac{1}{8 Z}(4Z+2(Z-2CM\braket{M|C}))\nonumber\\
&=&\frac{1}{4 Z}(2Z+(Z-2CM\braket{M|C}))\nonumber\\
&=&\frac{1}{4 Z}(3Z-2CM\braket{M|C})\nonumber\\
\implies\braket{M|C}&=&\frac{(3-4P_0)Z}{2CM}
\label{}
\end{eqnarray} 

The above equation \ref{inner product of vectors} states that after measurement, from the probability of 0, we obtain the inner product between the centroid and minority. In a slightly different perspective, let us calculate inner product of $\psi$ and $\phi$,

\begin{eqnarray}
\langle\phi \mid \psi\rangle=\left(\frac{\langle C| \otimes\langle 0|-\langle M| \otimes\langle 1|}{\sqrt{Z}}\right)\left(\frac{|0\rangle \otimes|C\rangle+|1\rangle \otimes|M\rangle}{\sqrt{2}}\right)
\end{eqnarray}

Expanding the inner product:
\begin{eqnarray}
\langle\phi \mid \psi\rangle&=&\frac{1}{\sqrt{Z}} \frac{1}{\sqrt{2}}(\langle C| \otimes\langle 0|(|0\rangle \otimes|C\rangle)+\langle C| \otimes\langle 0|(|1\rangle \otimes|M\rangle)\nonumber\\
&-&\langle M| \otimes\langle 1|(|0\rangle \otimes|C\rangle)-\langle M| \otimes\langle 1|(|1\rangle \otimes|M\rangle))
\end{eqnarray}

Simplifying each term:
\begin{eqnarray}
&&\langle C| \otimes\left\langle\left. 0|(|0\rangle \otimes|C\rangle)=\langle C \mid C\rangle \otimes\langle 0 \mid 0\rangle=| C\right|^2\right. \nonumber \\
&&\langle C| \otimes\langle 0|(|1\rangle \otimes|M\rangle)=0  \nonumber \\ 
&&\langle M| \otimes\langle 1|(|0\rangle \otimes|C\rangle)=0 \nonumber \\
&&\langle M| \otimes\left\langle\left. 1|(|1\rangle \otimes|M\rangle)=\langle M \mid M\rangle \otimes\langle 1 \mid 1\rangle=| M\right|^2\right.
\end{eqnarray}

So, the inner product simplifies to:
\begin{eqnarray}
\langle\phi \mid \psi\rangle&=&\frac{1}{\sqrt{Z}} \frac{1}{\sqrt{2}}\left(|C|^2-|M|^2\right) \nonumber \\
\langle\phi \mid \psi\rangle&=&\frac{|C|^2-|M|^2}{\sqrt{2 Z}}
\end{eqnarray}
Calculating $|\langle\phi \mid \psi\rangle|^2$ :
\begin{eqnarray}
|\langle\phi \mid \psi\rangle|^2=\left(\frac{|C|^2-|M|^2}{\sqrt{2 Z}}\right)^2=\frac{\left(|C|^2-|M|^2\right)^2}{2 Z}
\end{eqnarray}

\begin{eqnarray}
{2 Z}|\langle\phi \mid \psi\rangle|^2=2 Z\left(\frac{\left(|C|^2-|M|^2\right)^2}{2 Z}\right)
\end{eqnarray}

simplifying:
\begin{eqnarray}
{2 Z}|\langle\phi \mid \psi\rangle|^2=\left(|C|^2-|M|^2\right)^2
\end{eqnarray}

Assuming 
\begin{eqnarray}
{2 Z}|\langle\phi \mid \psi\rangle|^2 &=& D^2\nonumber\\
\implies D^2&=&2 Z|\langle\phi \mid \psi\rangle|^2
\end{eqnarray}
The term $D$ refers to the euclidean distance \cite{Mart_Dissimilarity_2023}, and the inner product of $\langle\phi \mid \psi\rangle$ represents the swaptest probability.

\begin{figure}[H]
\centering
\Qcircuit @C=4em @R=1.2em {
  \lstick{\ket{0}} & \gate{H} & \ctrl{2} & \gate{H} & \meter & \qw \\
  \lstick{q_1} & \gate{\ket{\phi}} & \qswap & \qw & \qw & \qw \\
  \lstick{q_2} & \multigate{5}{\ket{\psi}} & \qswap & \qw & \qw & \qw \\
  \lstick{q_3} & \ghost{\ket{\psi}} & \qw & \qw & \qw & \qw \\
  \lstick{q_4} & \ghost{\ket{\psi}} & \qw & \qw & \qw & \qw \\
  \lstick{q_5} & \ghost{\ket{\psi}} & \qw & \qw & \qw & \qw \\
  \lstick{q_6} & \ghost{\ket{\psi}} & \qw & \qw & \qw & \qw \\
  \lstick{q_7} & \ghost{\ket{\psi}} & \qw & \qw & \qw & \qw \\
  \lstick{C} & \cw & \cw & \cw & \cw \cwx[-8] & \cw 
}
\caption{Compact Swap test circuit.}
\label{Compact Swap test circuit.}
\end{figure}
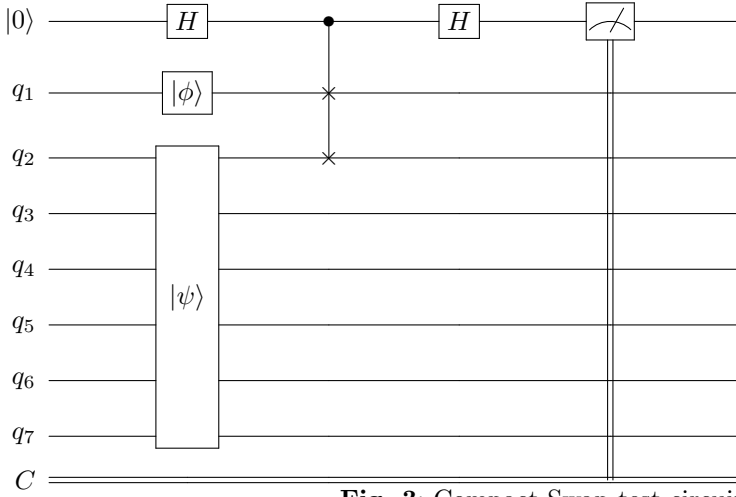

Based on this, we define the angle between two vectors or angular distance as 
\begin{eqnarray}
\text{angular\_distance} = 2 \cos^{-1}(\sqrt{\text{swap\_test\_probability}})
\end{eqnarray}
The above angular distance or the angle between two vectors will be used to rotate the minority class data point, which we will describe subsequently.

\begin{figure}[H]
\centering
\includegraphics[width=\linewidth]{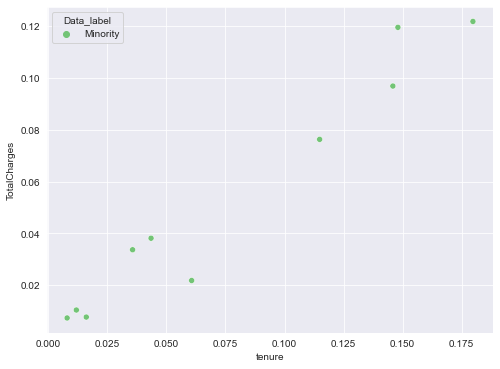}
\caption{Plot illustrating Sample data points of Minority class from population without any rotation.}
\label{Fig2_original}
\end{figure}

\begin{figure}[H]
\centering
\begin{subfigure}{0.33\linewidth}
\includegraphics[width=\linewidth]{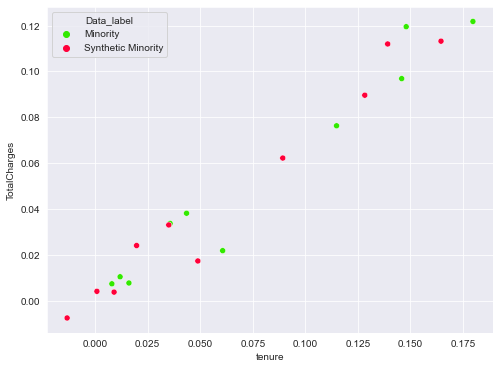} 
\caption{}
\label{qsmote_Fig2a}
\end{subfigure}\hfill
\begin{subfigure}{0.33\linewidth}
\includegraphics[width=\linewidth]{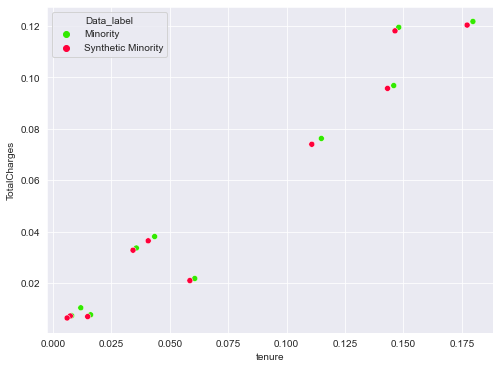} 
\caption{}
\label{qsmote_Fig2b}
\end{subfigure}\hfill
\begin{subfigure}{0.33\linewidth}
\includegraphics[width=\linewidth]{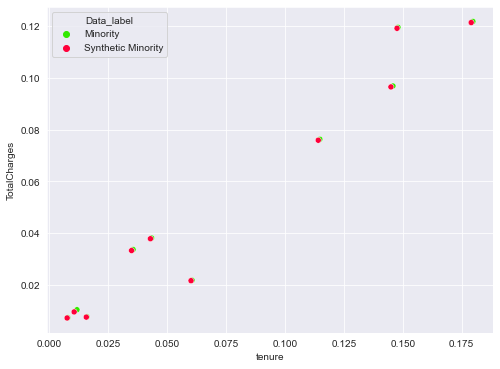} 
\caption{}
\label{qdctc_Fig2c}
\end{subfigure}\hfill
\caption{Plot illustrating impact of X Rotation on Sample data points of Minority class. (a) X Rotation with $split\_factor=2$, (b) X Rotation with $split\_factor=5$, (c) X Rotation with $split\_factor=10$.}
\label{qsmote_Fig3}
\end{figure}

\begin{figure}[H]
\centering
\begin{subfigure}{0.5\linewidth}
\includegraphics[width=\linewidth]{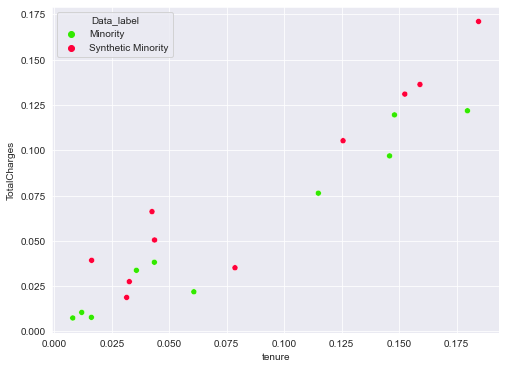} 
\caption{}
\label{qsmote_Fig3a}
\end{subfigure}\hfill
\begin{subfigure}{0.5\linewidth}
\includegraphics[width=\linewidth]{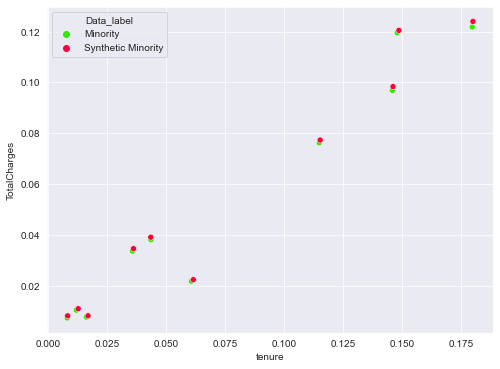} 
\caption{}
\label{qsmote_Fig3b}
\end{subfigure}\hfill
\caption{Plot illustrating impact of Y Rotation on Sample data points of Minority class. (a) Y Rotation with $split\_factor=5$, (b) Y Rotation with $split\_factor=100$}
\label{qsmote_Fig4}
\end{figure}

\begin{figure}[H]
\centering
\begin{subfigure}{0.5\linewidth}
\includegraphics[width=\linewidth]{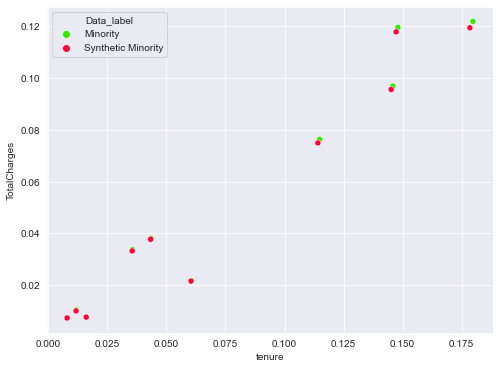} 
\caption{}
\label{qsmote_Fig4a}
\end{subfigure}\hfill
\begin{subfigure}{0.5\linewidth}
\includegraphics[width=\linewidth]{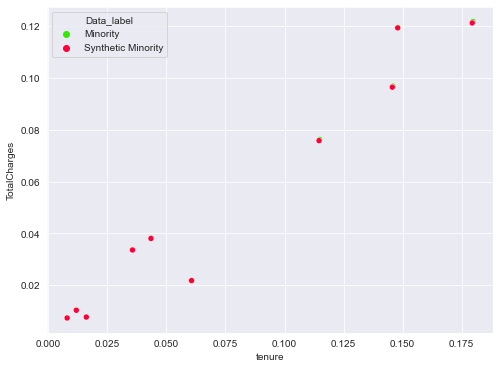} 
\caption{}
\label{qsmote_Fig4b}
\end{subfigure}\hfill
\caption{Plot illustrating impact of Z Rotation on Sample data points of Minority class. (a) Z Rotation with $split\_factor=5$, (b) Z Rotation with $split\_factor=10$}
\label{qsmote_Fig5}
\end{figure}

\subsection{Applying Rotation to data point} \label{Methodology Rotation}
After calculating the angle (angular distance) between two vectors, we rotate the actual minority data point by an angle less than the calculated angle to create a synthetic data point. We choose to minimize the angle of rotation to prevent abrupt fluctuation of values in the minority data point. We perform rotations along the X, Y, and Z axes to analyze their impact on the minority data points.

As the angle of rotation is minimal for the minority data point vector, we have derived the angle of rotation with the below logic. \\
\begin{algorithm}
\caption{Angle of rotation calculation logic}
\begin{algorithmic}
\State \textbf{sf}: split\_factor
\If{$\text{angular\_distance} > \frac{\pi}{2}$}
    \State $\text{angle} \gets \left| \frac{\pi}{2} - \text{angular\_distance} \right| / \text{sf}$
\ElsIf{$\text{angular\_distance} < 0$}
    \State $\text{angle} \gets \left| \left( \frac{\pi}{2} - \text{angular\_distance} \right) \times \text{random}(0.5, 1) \right| / \text{sf}$
\Else
    \State $\text{angle} \gets \text{random}(0, \text{angular\_distance}) / \text{sf}$
\EndIf
\end{algorithmic}
\end{algorithm}

$split\_factor$ is the factor by which we want to divide the generated angle, we have experimented with 2, 5, 10 and 100 for various rotations mentioned above and will outline the result of rotation for a sample containing 10 data points.

The aforementioned figures illustrate the influence of rotation on data points. Initially, we selected a subset of data points from the minority class and visually represented them in Figure  \ref{Fig2_original}. This figure displays a scatter plot of the data points. Subsequently, we have performed X, Y, and Z rotations on these data points, using the $split\_factor$ as a basis. We conducted experiments by incrementally increasing the split factor and evaluating the resulting effects on rotations.

\textbf{X Rotation}: X Rotation refers to the rotation of a data point in relation to the X axis. We conducted experiments with split factors of 2, 5, and 10.  Upon increasing the split factor from 2 to 10, we see that the synthetic data points created by each split factor exhibit a greater proximity to the original data points. When the split factor 2 is used for X rotation, the resulting data points are located at a certain distance from the original location. As we go from 5 to 10, the freshly created data points get increasingly closer together. At 10, the synthetic data point is the closest among the three dividing factors. 

\textbf{Y Rotation}: Y Rotation refers to the rotational movement of data points around the Y axis. From the analysis of figure \ref{qsmote_Fig4}, it is evident that the newly created data points exhibit a high sensitivity to Y rotations. Additionally, these data points need the generation of extremely tiny angles in order to be positioned in close proximity to the Source(the minority sample). It is evident that as the splitting factor ($100$) increases, resulting in extremely tiny angles, the created data point is closest to the source. Conversely, small splitting factors ($5$) yield data points that deviate significantly from the nature of the data point sample.

\textbf{Z Rotation}: Z rotation refers to the rotation of data points around the Z axis. Based on the evidence shown in Figure \ref{qsmote_Fig5}, we can confidently infer that the behavior of Z rotation is similar to that of X rotation. Additionally, it is evident that using splitting factors of 5 and 10 results in the generation of additional data points that are in close proximity to the source.

In general, it can be confidently said that all rotations have the ability to generate synthetic data points. However, the Y rotation is more sensitive, but the X and Z rotations provide similar outcomes.

\subsection{Quantum SMOTE Algorithm} \label{Approach}
We now introduce QuantumSMOTE. Broadly, our algorithm proceeds in two steps: clustering of the population and generating synthetic data points by the swap test and rotation of minority class data points. We believe clustering is an essential pre-step to synthetic data generation. Though we can use any clustering method that produces clusters in data, we have used K-Means Clustering in our research with a minimum of 3 clusters, and we recommend the same for further research on this topic. 

Post clustering, we proceed with synthetic data generation, and for the purpose of simplicity, we name this part the QuantumSMOTE function. The pseudocode of this is described in the section below. Generally, it comprises four distinct parts: Data preparation for the swap test, application of the swap test, rotation of synthetic data points, and generation of synthetic data points for each cluster based on the target.

\begin{algorithm*}
\caption{Preparation for Swap Test}
\label{algo_swap test_prep}
\begin{algorithmic}[1]
\Function{Prepswap test}{$data\_point1$, $data\_point2$}
    \State $norm\_data\_point1 \gets 0$
    \State $norm\_data\_point2 \gets 0$
    \State $Dist \gets 0$
    \For{$i \gets 0$ \textbf{to} $length(data\_point1)-1$}
        \State $norm\_data\_point1 \gets norm\_data\_point1 + data\_point1[i]^2$
        \State $norm\_data\_point2 \gets norm\_data\_point2 + data\_point2[i]^2$
        \State $Dist \gets Dist + (data\_point1[i] + data\_point2[i])^2$
    \EndFor
    \State $Dist \gets \sqrt{Dist}$
    \State $data\_point1\_norm \gets \sqrt{norm\_data\_point1}$
    \State $data\_point2\_norm \gets \sqrt{norm\_data\_point2}$
    \State $Z \gets \text{round}(data\_point1\_norm^2 + data\_point2\_norm^2)$
    \State $\phi \gets [\text{round}(data\_point1\_norm / \sqrt{Z}, 3), -\text{round}(data\_point2\_norm / \sqrt{Z}, 3)]$
    \State Initialize array $\psi$
    \For{$i \gets 0$ \textbf{to} $length(data\_point1)-1$}
        \State $\psi.\text{append}(\text{round}(data\_point1[i] / (data\_point1\_norm \times \sqrt{2}), 3))$
        \State $\psi.\text{append}(\text{round}(data\_point2[i] / (data\_point2\_norm \times \sqrt{2}), 3))$
    \EndFor
    \State \Return $\phi, \psi$
\EndFunction
\end{algorithmic}
\end{algorithm*}

\begin{algorithm}
\caption{Swap Test }
\label{algo_swap test}
\begin{algorithmic}[1]
\Function{swap testV1}{$\psi$, $\phi$}
    \State Initialize Quantum Register $q1$ with 1 qubit
    \State Initialize Quantum Register $q2$ with n+2 qubits
    \State Initialize Classical Register $c$ with 1 bit
    \State Create Quantum Circuit with $q1$, $q2$, and $c$

    \textbf{States initialization}
    \State Initialize $q2[0]$ with state $\phi$
    \State Initialize $q2[1:n+2]$ with state $\psi$

    \textbf{The swap test operator}
    \State Apply Pauli-X Gate to $q2[1]$

    \textbf{Swap Test}
    \State Apply Hadamard Gate to $q1[0]$
    \State Apply Controlled SWAP Gate on $q1[0]$, $q2[0]$, and $q2[1]$
    \State Apply Hadamard Gate to $q1[0]$
    \State Measure $q1$ into classical register $c$

    \textbf{Simulation and result collection}
    \State Set up quantum simulator
    \State Execute the quantum circuit on the simulator
    \State Collect the result into a variable $result$
    \State Extract measurement counts from $result$

    \textbf{Calculate the Swap Test probability}
    \State $p0 \gets \frac{\text{counts.get('0', 0)}}{total\_shots}$
    \State $p1 \gets \frac{\text{counts.get('1', 0)}}{total\_shots}$
    \State $swap\_test\_probability \gets 1 - 2 \times p0 + p1$
    \State Print $swap\_test\_probability$

    \textbf{Calculate the angular distance}
    \State $angular\_distance \gets 2 \times \text{arccos}(\sqrt{swap\_test\_probability})$
    \State Print $angular\_distance$

    \State \Return $swap\_test\_probability$, $angular\_distance$
\EndFunction
\end{algorithmic}
\end{algorithm}

\begin{algorithm}
\caption{Normalize Array}
\label{algo_Normalize_Array}
\begin{algorithmic}[1]
\Function{NormalizeArray}{$arr$}
    \textbf{Calculate the sum of squares of the elements in the array}
    \State $sum\_of\_squares \gets \Call{SumOfSquares}{arr}$

    \textbf{Check if the sum of squares is already very close to 1}
    \If{\Call{IsClose}{$sum\_of\_squares$, $1.0$, $\text{rtol}=1e-6$}}
        \State \Return $arr$
    \EndIf

    \textbf{Calculate the scaling factor to make the sum of squares equal to 1}
    \State $scaling\_factor \gets 1.0 / \sqrt{sum\_of\_squares}$

    \textbf{Normalize the array by multiplying each element by the scaling factor}
    \State $normalized\_arr \gets arr \times scaling\_factor$

    \State \Return $normalized\_arr$
\EndFunction
\end{algorithmic}
\end{algorithm}

\begin{algorithm}
\caption{Create Synthetic Data}
\label{algo_synthetic}
\begin{algorithmic}[1]
\State \textbf{$ad:$} $angular\_distance$
\State \textbf{$sf:$} $split\_factor$
\Function{CreateSynData}{$n$, $loop\_ctr$, $angle\_increment$, $ad$,$sf$, $data\_point1$, $data\_point2$}
    \State $data\_point1 \gets \Call{NormalizeArray}{data\_point1}$
    \State $data\_point2 \gets \Call{NormalizeArray}{data\_point2}$
    \State Initialize Quantum Circuit $circuit$ with $n$ qubits
    \State $circuit.\Call{Initialize}{data\_point1}$

    \If{$ad > \frac{\pi}{2}$}
        \State $angle \gets \left| \frac{\pi}{2} - angular\_distance \right| / sf$
    \ElsIf{$ad < 0$}
        \State $angle \gets \left| \frac{\pi}{2} - ad \right| \times \Call{RandomUniform}{0.5, 1}/sf$
    \Else
        \State $angle \gets \Call{RandomUniform}{0, ad} / sf$
    \EndIf

    \State Print "rotation angle", $angle$
    \State $angle \gets angle + angle\_increment$

    \For{$l \gets 0$ \textbf{to} $n-1$}
        \State Apply RX gate to $circuit$ at qubit $l$ with angle $angle$
    \EndFor

    \textbf{Simulate the quantum circuit}
    \State Set up quantum simulator
    \State Execute $circuit$ on the simulator and store result in $job$
    \State $result \gets job.result()$
    \State $statevector \gets result.get\_statevector()$

    \textbf{Extract the final data point from the statevector}
    \State $new\_data\_point \gets \Call{Real}{statevector}$

    \State \Return $new\_data\_point$
\EndFunction
\end{algorithmic}
\end{algorithm}

\begin{algorithm}
\caption{Quantum Synthetic Minority Over-sampling Technique}
\begin{algorithmic}[1]
\Function{QuantumSMOTE}{Data, Target\_pct, cluster\_centroids}
    \State Create an empty DataFrame $syn\_dataframe$
    \State $target\_synthetic\_percent \gets 30$

    \For{each cluster with index $clus\_idx$ in $centroid\_df$}
        \State $minority\_count\_in\_cluster \gets$ Find number of minority samples in the cluster
        \State $total\_count\_in\_cluster \gets$ Find total number of samples in the cluster
        \State $minority\_percent \gets$ Calculate minority percentage in the cluster
        \State Print "minority\% in cluster $clus\_idx$ is =", $minority\_percent$

        \State $synthetic\_loop\_itr \gets (\text{Target\_pct} - minority\_percent) / minority\_percent$
        \State Print "Number of synthetic datapoint iteration is =", $synthetic\_loop\_itr$

        \If{$synthetic\_loop\_itr > 0$ and $synthetic\_loop\_itr < 1$}
            \State $synthetic\_loop\_itr1 \gets 1$
        \ElsIf{$synthetic\_loop\_itr > 1$}
            \State $synthetic\_loop\_itr1 \gets \Call{Ceil}{synthetic\_loop\_itr}$
            \State $fraction\_part \gets synthetic\_loop\_itr1 - \Call{Floor}{synthetic\_loop\_itr}$
        \Else
            \State $synthetic\_loop\_itr1 \gets -1$
        \EndIf

        \If{$synthetic\_loop\_itr1 \geq 0$}
            \For{$syn\_loop \gets 0$ \textbf{to} $synthetic\_loop\_itr1 - 1$}
                \If{$syn\_loop = synthetic\_loop\_itr1 - 1$}
                    \State Select $centroid\_temp$ and $minority\_temp$ as a fraction of minority in cluster $clus\_idx$
                \Else
                    \State Select $centroid\_temp$ and $minority\_temp$ as the entire minority data for cluster $clus\_idx$
                \EndIf

                \State Flatten $centroid\_temp$ to $centroid\_dp\_tmp$
                \For{each row in $minority\_temp$}
                    \State Select $minority\_dp\_temp$ as the current row
                    \State Calculate $phi$ and $psi$ using \Call{prep\_swap test}{$minority\_dp\_temp$, $centroid\_dp\_tmp$}
                    \State Normalize $phi$ and $psi$ to $phi1$ and $psi1$
                    \State Calculate $swap\_test\_probability$ and $angular\_distance$ using \Call{swap test\_v1}{$psi1$, $phi1$}
                    \State $n \gets \Call{LogBase2}{\text{length of} minority\_dp\_temp}$

                    \If{length of $minority\_dp\_temp$ is not divisible by $n$}
                        \State $add \gets 1$
                    \Else
                        \State $add \gets 0$
                    \EndIf

                    \State $loop\_ctr \gets \Call{Round}{\text{length of} minority\_dp\_temp / n + add}$
                    \State $angle\_increment \gets syn\_loop \times 0.0174533$
                    \State $syn\_data \gets$ \Call{create\_syn\_data}{$n$, $loop\_ctr$, $angle\_increment$, $angular\_distance$, $minority\_dp\_temp$, $centroid\_dp\_tmp$}
                    \State Create DataFrame $syn\_df\_temp$ from $syn\_data$
                    \State Concatenate $syn\_df\_temp$ with $syn\_dataframe$
                \EndFor
            \EndFor
        \ElsIf{$synthetic\_loop\_itr1 < 0$}
            \State Print "Cluster $clus\_idx$ has already a high percentage of minority $minority\_percent$. Close to target synthetic percent $target\_synthetic\_percent$."
        \Else
            \State Print "Nothing to process..."
        \EndIf
    \EndFor
    \State \Return $syn\_dataframe$
\EndFunction
\end{algorithmic}
\label{Algo_Qsmote}
\end{algorithm}

\section{Case Study and Results} \label{Case Study and Results}
To test the QuantumSMOTE algorithm, we analyse the publicly available dataset of telecom churn \cite{telco_churn_dataset_kaggle}. This dataset is widely used to experiment and test various models for customer retention and is quite useful in comparing classical models with the models post-induction of synthetic data by the quantum SMOTE algorithm. In the following subsections, we will describe data behavior, data preparation for modeling, and applying QuantumSMOTE on the data.

\subsection{Improving Telecom Churn Prediction Using SMOTE} \label{Improving Telecom Churn Prediction Using SMOTE}

The telecom churn dataset is purposefully developed to predict customer behavior and help in generating customer retention programs. Each row in the dataset represents an individual consumer, with each column representing different attributes of these customers. Notably, the dataset has such characteristics as:

\textbf{Churn Indicator}: This column identifies customers who have terminated their service during the previous month.

\textbf{Subscribed Services}: A detailed list of all services that each customer has signed up for, such as phone service, multiple lines, internet, online security, online backup, device protection, tech support, and streaming TV and movies.

\textbf{Account Information}:  comprises of how long they have been a client for, the terms of the contract they entered into with their company, and which method they would prefer to use when making payments so as to keep track of their spending habits effectively through electronic means like electronic mail that may save on transaction costs like envelope usage, monthly expenditure and cumulative costs incurred so far.

\textbf{Demographic Information}: It provides information about the customer's gender, age group, whether or not they are married, and whether they have dependent children.

\subsubsection{Preparing Data For Quantum SMOTE}
The Telco churn dataset is amenable to a usual data preparation process, which broadly includes the following phases.

\textbf{Missing Value Tearment}: Inspect the telco churn dataset for null values and adapt a strategy to handle them. Since we found a very small percentage of records ($11$, to be precise) that have missing values across multiple columns, we proceeded with dropping them.

\textbf{Removing Irrelevant Data}: Identify and remove any columns that are not relevant to churn prediction, such as customer IDs that are unique and not predictive of churn.

\textbf{Data Type Convertion}: To ensure that each column is of the appropriate data type, we have converted multiple columns with text data as to category. These included columns such as PhoneService, MultipleLines, InternetService, OnlineSecurity, OnlineBackup, DeviceProtection, TechSupport, StreamingTV, StreamingMovies, Contract, PaperlessBilling, PaymentMethod, gender, SeniorCitizen, Partner and Dependents. We also converted the numerical columns such as TotalCharges, tenure and MonthlyCharges to float to avoid any of them being treated as text due to import issues.

\textbf{Exploratory Data Analysis (EDA)}:
EDA was performed to understand the distribution and relationship of variables. We applied various univariate, and bivariate analyses to understand the behavior of data, particularly numeric variables, which are essential for creating models. The variables TotalCharges, tenure, and MonthlyCharges are particularly important since the distribution of these variables later will be used to verify the effect of the SMOTE procedure.

\textbf{Label Encoding}:
For the sake of better visualization and correlation analysis with the target variable, we performed label encoding of multiple categorical variables. 

\textbf{Correlation Analysis}:
We conducted a correlation analysis of numerical variables to eliminate multicollinearity. Also, we conducted a correlation analysis of all the variables with the target to select the best-fit variables for modeling. Post correlation analysis, we are able to drop multiple variables that are not relevant for the purpose of modeling.

\textbf{Onehot Encoding}:
Post selection of features, we converted all the categorical variables to Onehot encoding, thereby creating multiple numerical columns for each categorical value.

\textbf{Feature Scaling}:
Since Onehot encoding created multiple numerical columns with values 0 and 1, the continuous variables such as TotalCharges, tenure and MonthlyCharges are scaled by minmax scaling to lie between 0 and 1.

\subsubsection{Clustering}
As we have indicated earlier, the Quantum SMOTE algorithm relies on unique customer segments to calculate the angle between the segment centroid (mean) and minority data point; we have used the K-Means clustering method approach to derive segments. The approach for identifying inherent groupings among customers is based on their attributes, which can further assist in understanding customer behavior and improving retention strategies. For the sake of our experiment, we have identified 3 clusters using the K-Means approach to generate new data and highlight the achievements. The outcome of the clustering approach is at least 3 clusters (datasets that are dynamically segmented) with different majority-minority populations. These are useful when deriving angles based on which minority population across clusters will be most valuable for the SMOTE algorithm. 

\subsubsection{Quantum SMOTE and Synthetic Data}
After applying the Clustering algorithm to the Telecom Churn dataset and processing the data, we proceeded to apply the Quantum SMOTE Algorithm (\ref{Algo_Qsmote}) to each cluster. The goal was to enhance the representation of the minority population to a certain percentage of the overall dataset. The procedure used two primary approaches previously mentioned, namely the swap test (Algo. \ref{algo_swap test}) and rotation (Algo. \ref{algo_synthetic}). 

\textbf{Swap Test}: The fundamental operation of the swap test has been previously explained in the preceding sections. We use the swap test in a modified manner \ref{Compact Swaptest} to compute the angular distance between the vector representing the minority data point and the vector representing the centroid. The procedure is effectively executed in Ref. \cite{qiskit_medium, Mart_Dissimilarity_2023}. The swap test requires two inputs, denoted as $\phi$ and $\psi$. The state $\phi$ is determined by computing the norms of the inputs, which consist of the centroid and minority data points. On the other hand, the state $\psi$ is obtained by concatenating the normalized components of the inputs.  The execution of this preparation is shown in the auxiliary function \ref{algo_swap test_prep}. The circuit that is obtained is rendered in Fig. \ref{Compact Swap test circuit.}.

The main purpose of using this technique to swap test is to minimize the required number of qubits in constructing the swap test circuit, which becomes particularly advantageous as the dataset dimension expands. After performing feature selection and Onehot encoding, we obtained a final count of 32 columns. Consequently, our swap test circuit necessitates the use of 8 qubits and a classical register. Nevertheless, using a traditional methodology may have resulted in the use of 65 qubits. The swap test circuit facilitates the calculation of the angular distance between the cluster centroid and the minority data point.

\begin{figure}[H]
    \centering
    \includegraphics[width=\linewidth]{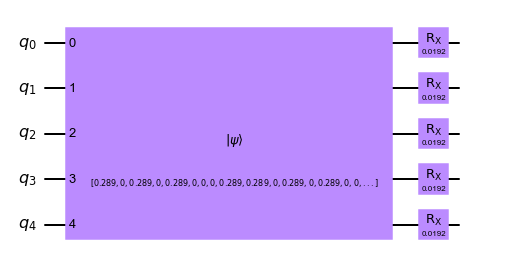}
    \caption{Data point rotation circuit.}
\label{qsmote_rotationcircuit}
\end{figure}

\begin{figure}[H]
\centering
\begin{subfigure}{0.5\linewidth}
\includegraphics[width=\linewidth]{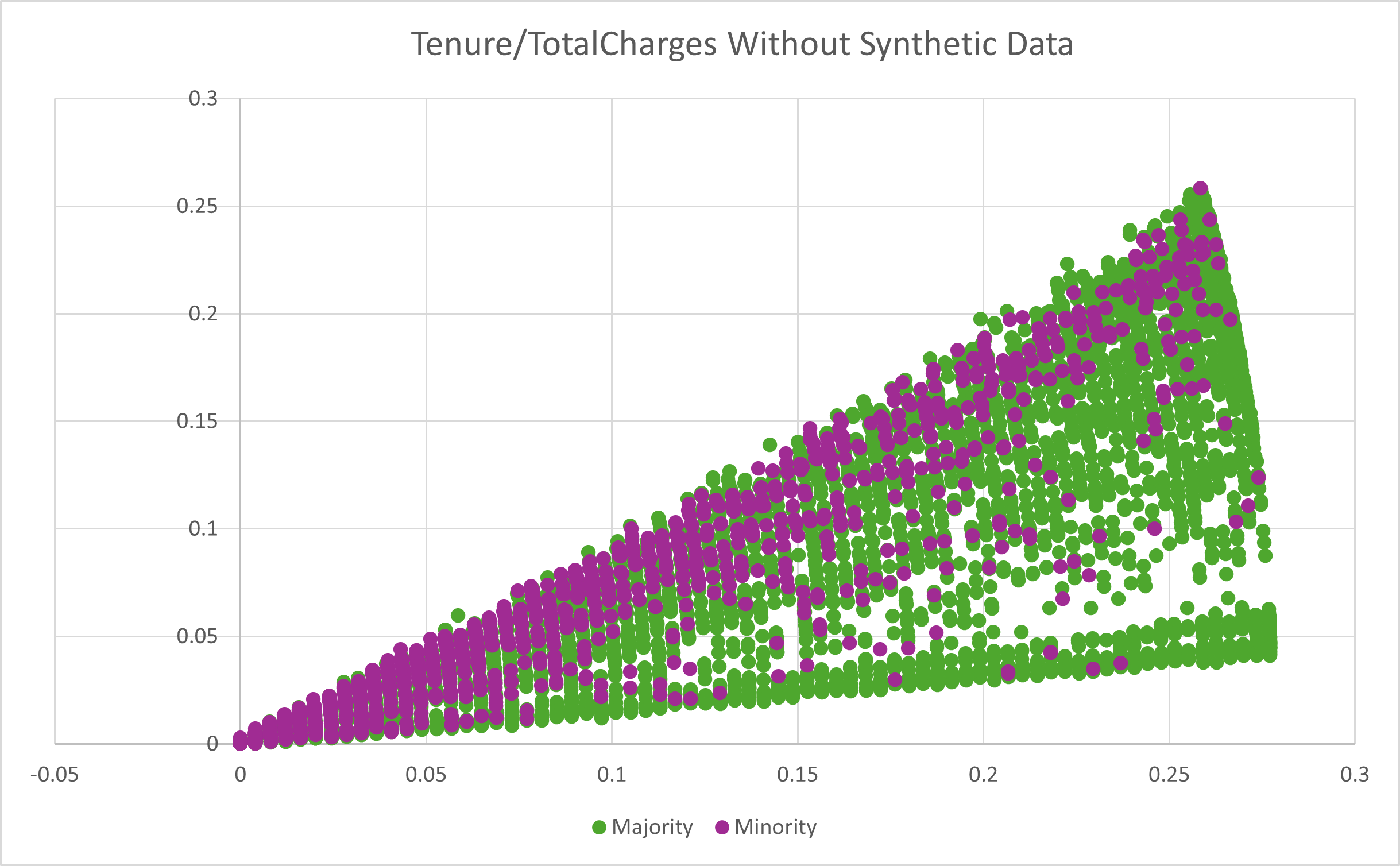} 
\caption{}
\label{qsmote_Fig45a}
\end{subfigure}\hfill
\begin{subfigure}{0.5\linewidth}
\includegraphics[width=\linewidth]{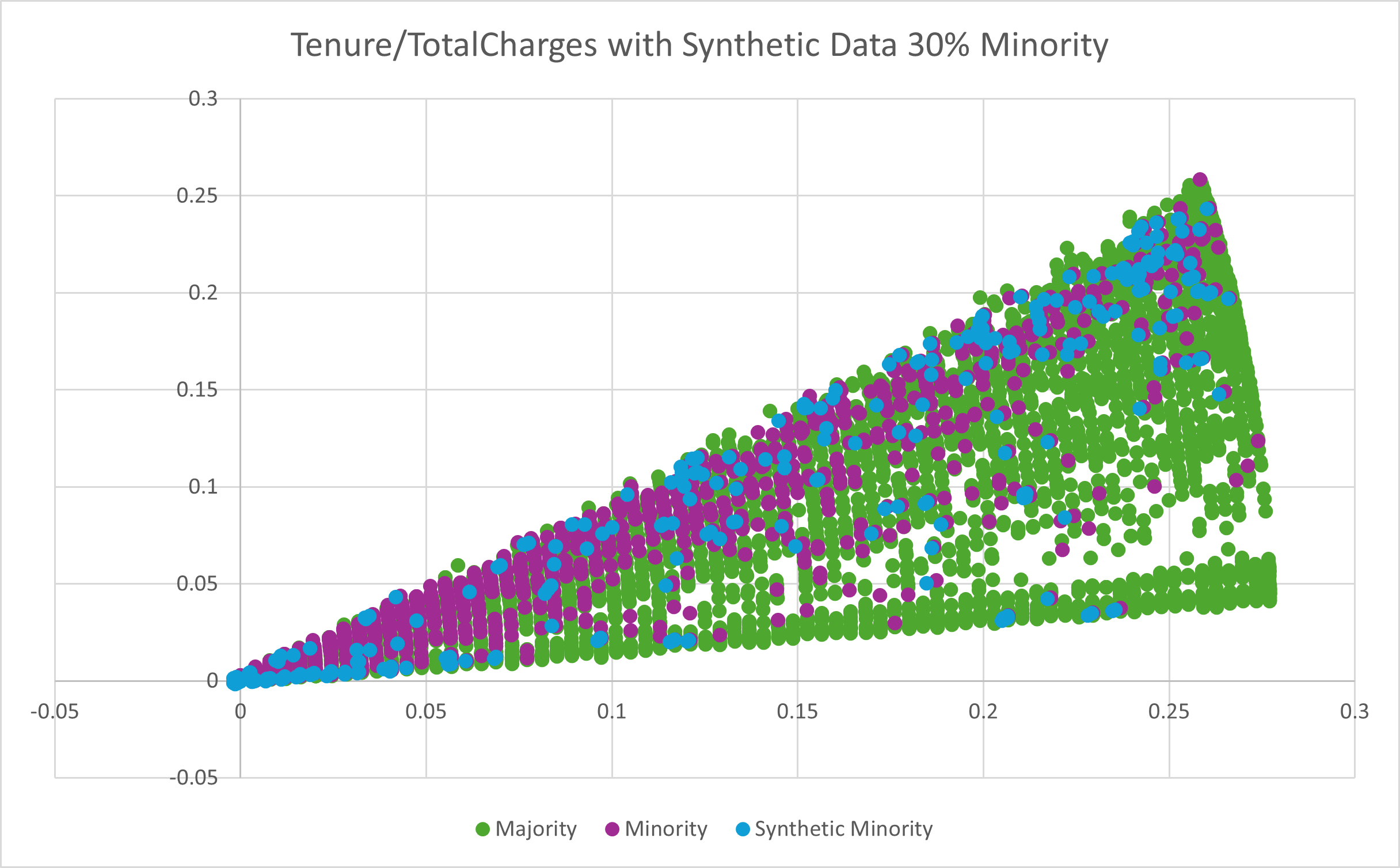} 
\caption{}
\label{qsmote_Fig45b}
\end{subfigure}\hfill
\begin{subfigure}{0.5\linewidth}
\includegraphics[width=\linewidth]{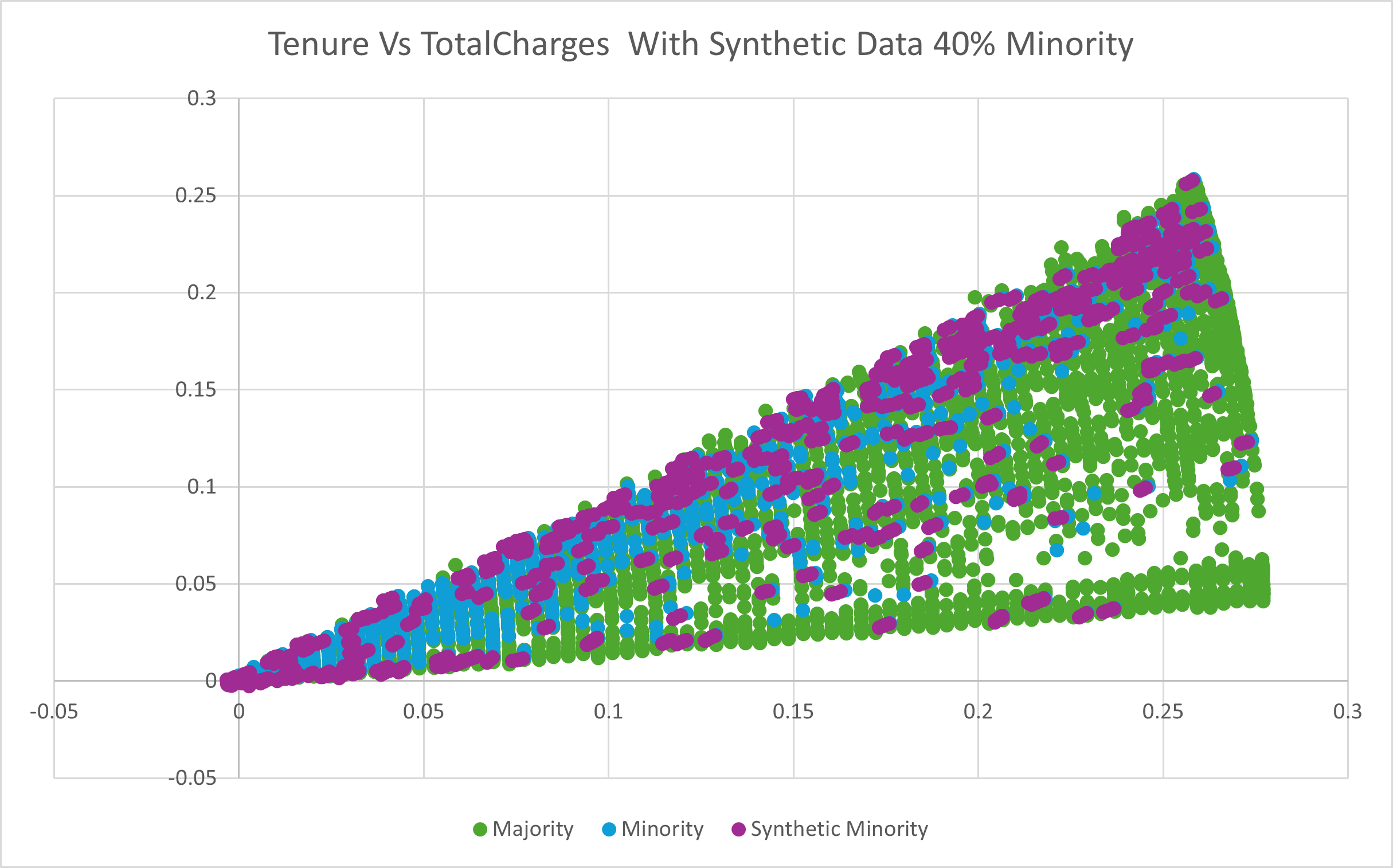} 
\caption{}
\label{qsmote_Fig45c}
\end{subfigure}\hfill
\begin{subfigure}{0.5\linewidth}
\includegraphics[width=\linewidth]{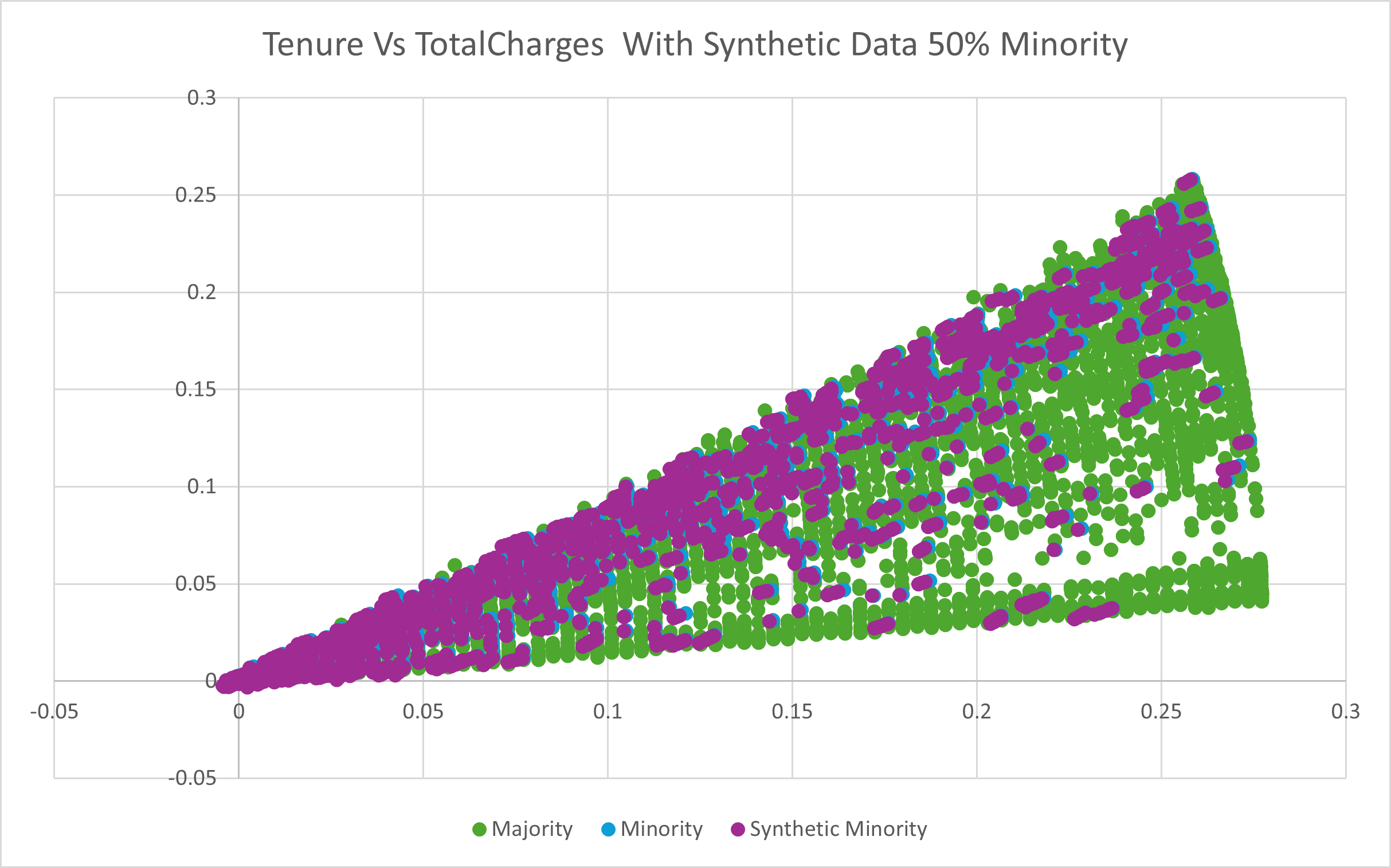} 
\caption{}
\label{qsmote_Fig45d}
\end{subfigure}\hfill
\caption{Plot illustrating impact of synthetic data generation on Sample data points of Minority class. (a) data points with no synthetic, (b) 30\% synthetic, (c) 40\% synthetic, (d) 50\% synthetic.}
\label{qsmote_Fig45}
\end{figure}

\textbf{Rotation}: After performing the swap test, it is necessary to rotate the minority data point by an angle that represents a minute fraction of the total angular distance. The rotation circuit executes the rotation of the normalized minority data point vector. In the preceding section \ref{Methodology Rotation}, we have provided a detailed explanation of the different rotations of X, Y, and Z. In this experiment, we applied X rotations to all of the minority data points. To account for numerous interactions or repeated rotations of a single minority data point, we have adjusted the rotation angle by $0.0174$, which corresponds to the conversion from radians to degrees. We are attempting to adjust the angle of the minority data point using angular degrees, even though the angular distance generated by the swap test is in radians. The rotation circuit comprises the state vector of the normalized data point and rotation gates (Fig. \ref{qsmote_rotationcircuit}).
By rotating minority data points, synthetic data points that closely resemble the original data points are created, thanks to the use of modest rotation angles. When the synthetic data points are included in the original dataset, it leads to an increase in the total density of the minority class. The scatter distribution of synthetic data points in the population is shown in Fig. \ref{qsmote_Fig45}. The data illustrates the distribution of classes (majority, minority, and synthetic minority) as the proportion of the minority class increases from $30\%$ to $50\%$.

\begin{figure}[H]
\centering
\begin{subfigure}{0.33\linewidth}
\includegraphics[width=\linewidth]{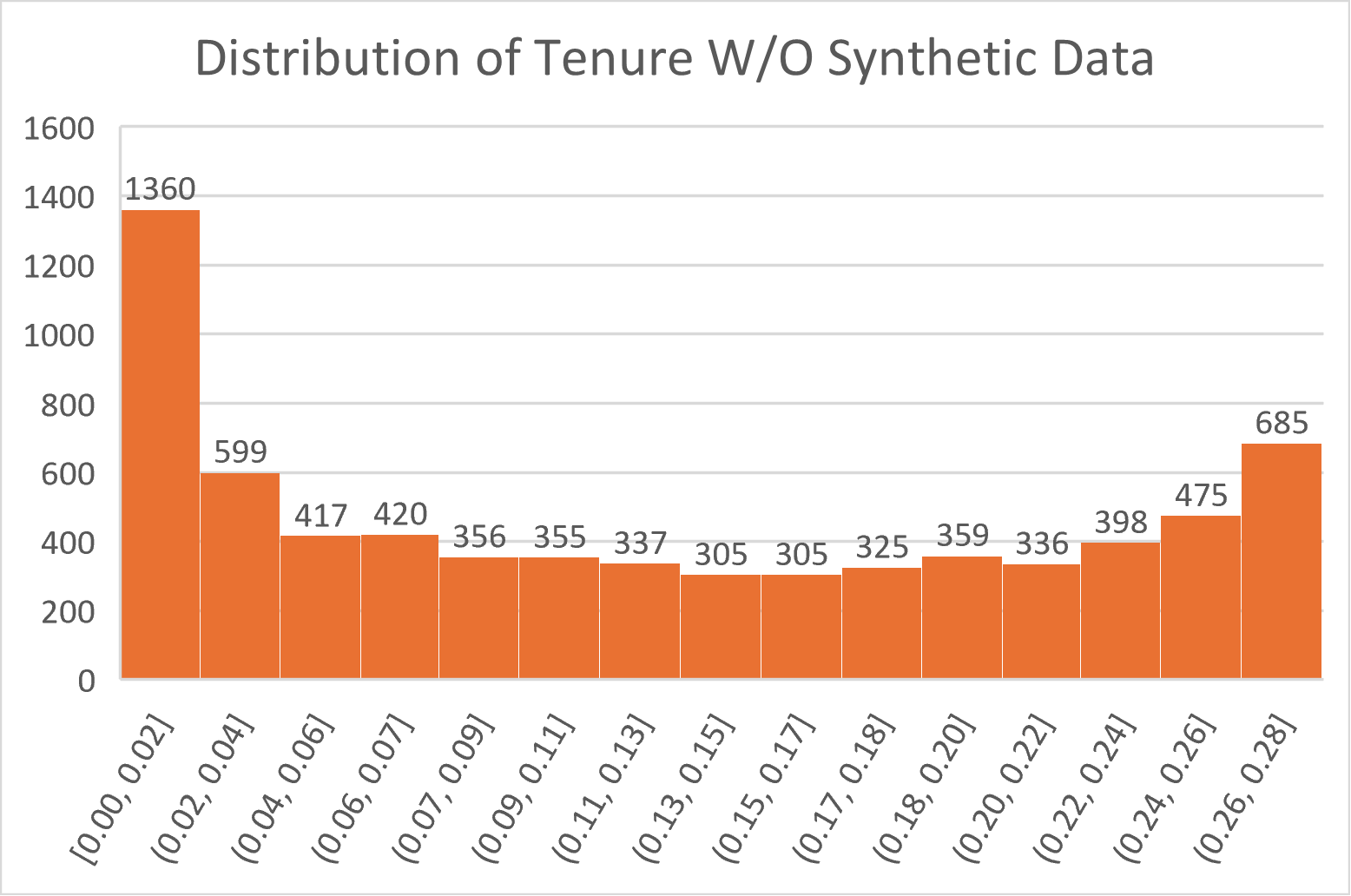} 
\caption{}
\label{qsmote_Fig6a}
\end{subfigure}\hfill
\begin{subfigure}{0.33\linewidth}
\includegraphics[width=\linewidth]{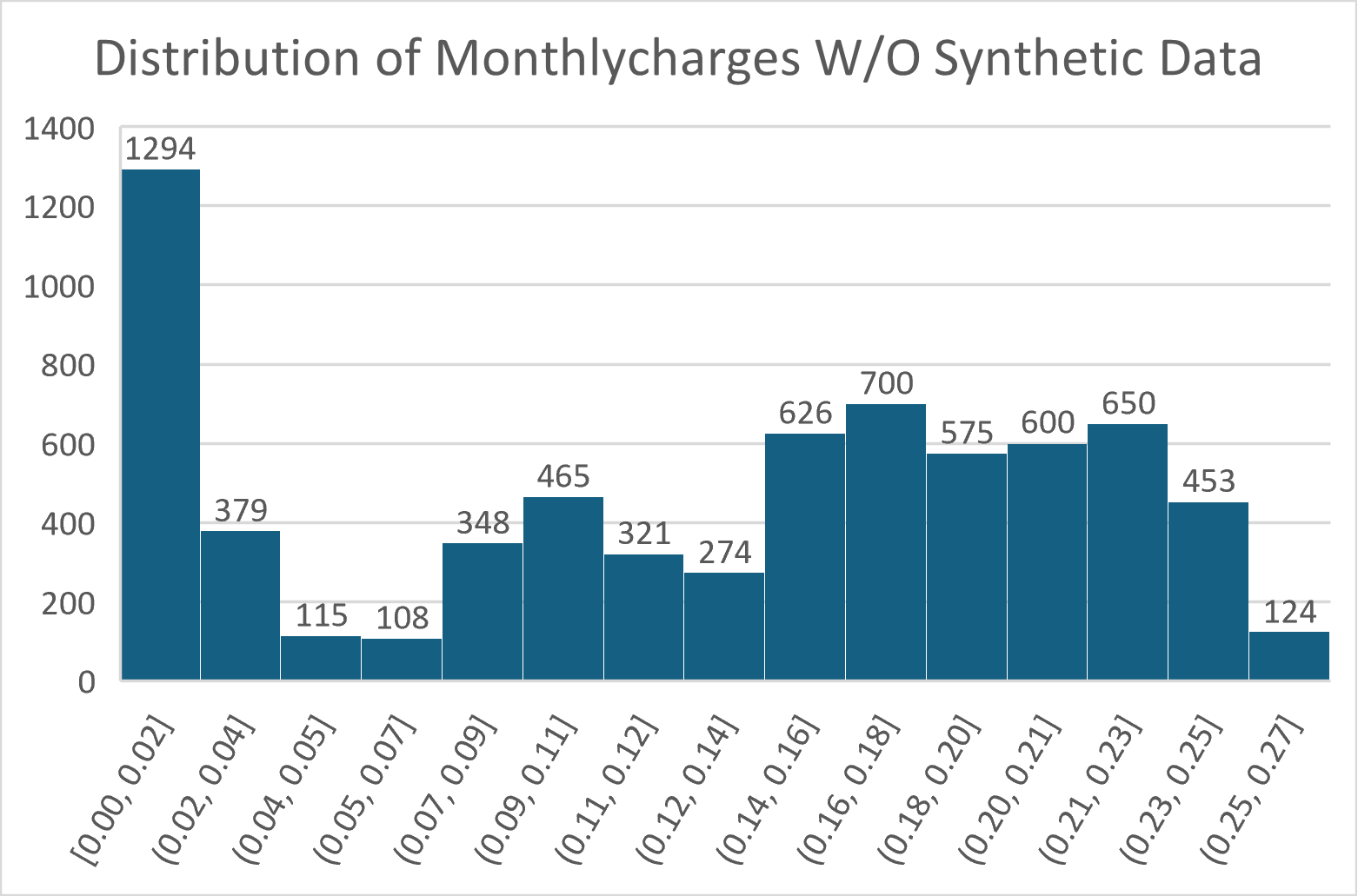} 
\caption{}
\label{qsmote_Fig6b}
\end{subfigure}\hfill
\begin{subfigure}{0.33\linewidth}
\includegraphics[width=\linewidth]{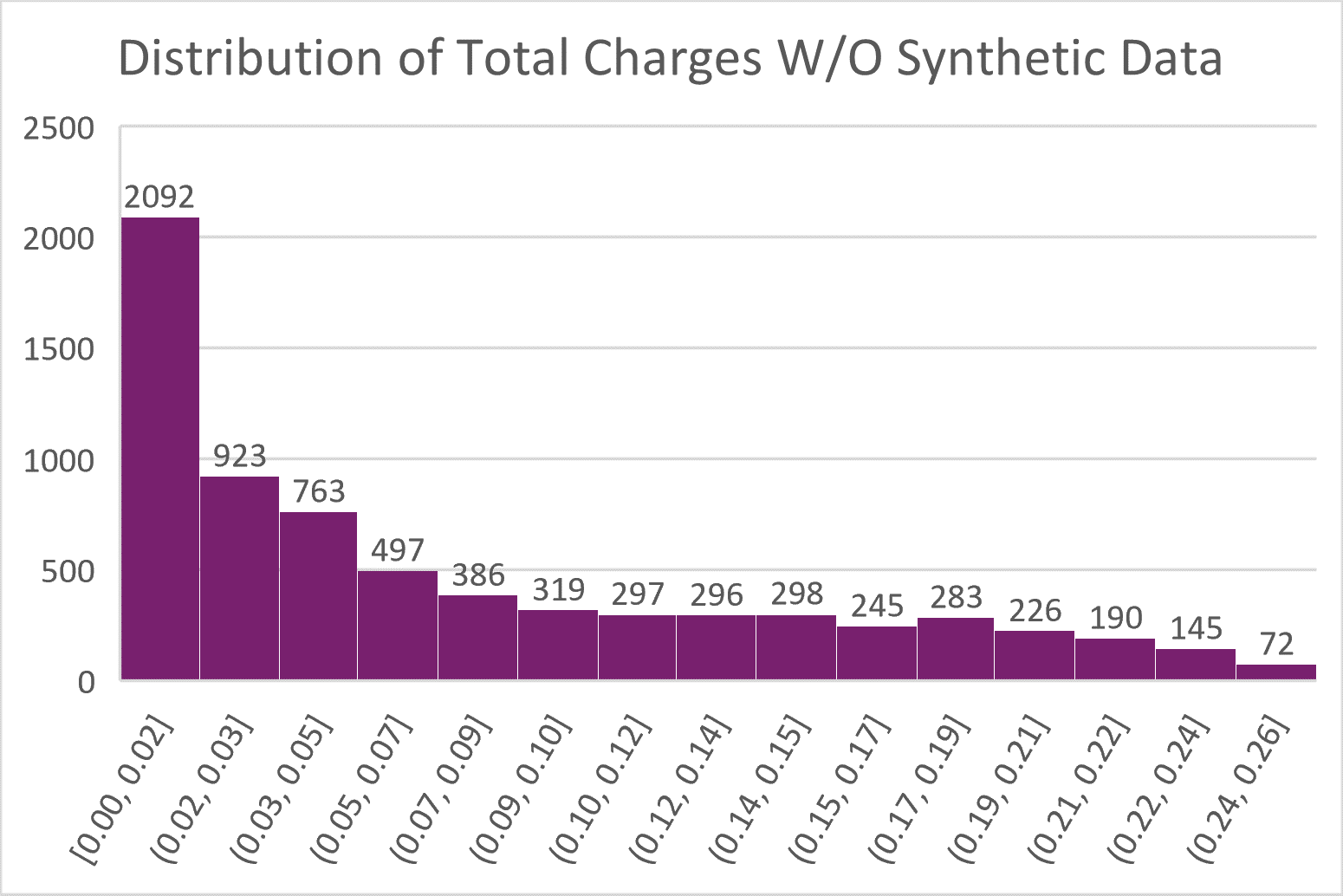} 
\caption{}
\label{qdctc_Fig6c}
\end{subfigure}\hfill
\caption{Plot illustrating distribution of 3 columns: (a) Tenure, (b) MonthlyCharges, and (c) TotalCharges.}
\label{qsmote_Fig7}
\end{figure}

\begin{figure}[H]
\centering
\begin{subfigure}{0.33\linewidth}
\includegraphics[width=\linewidth]{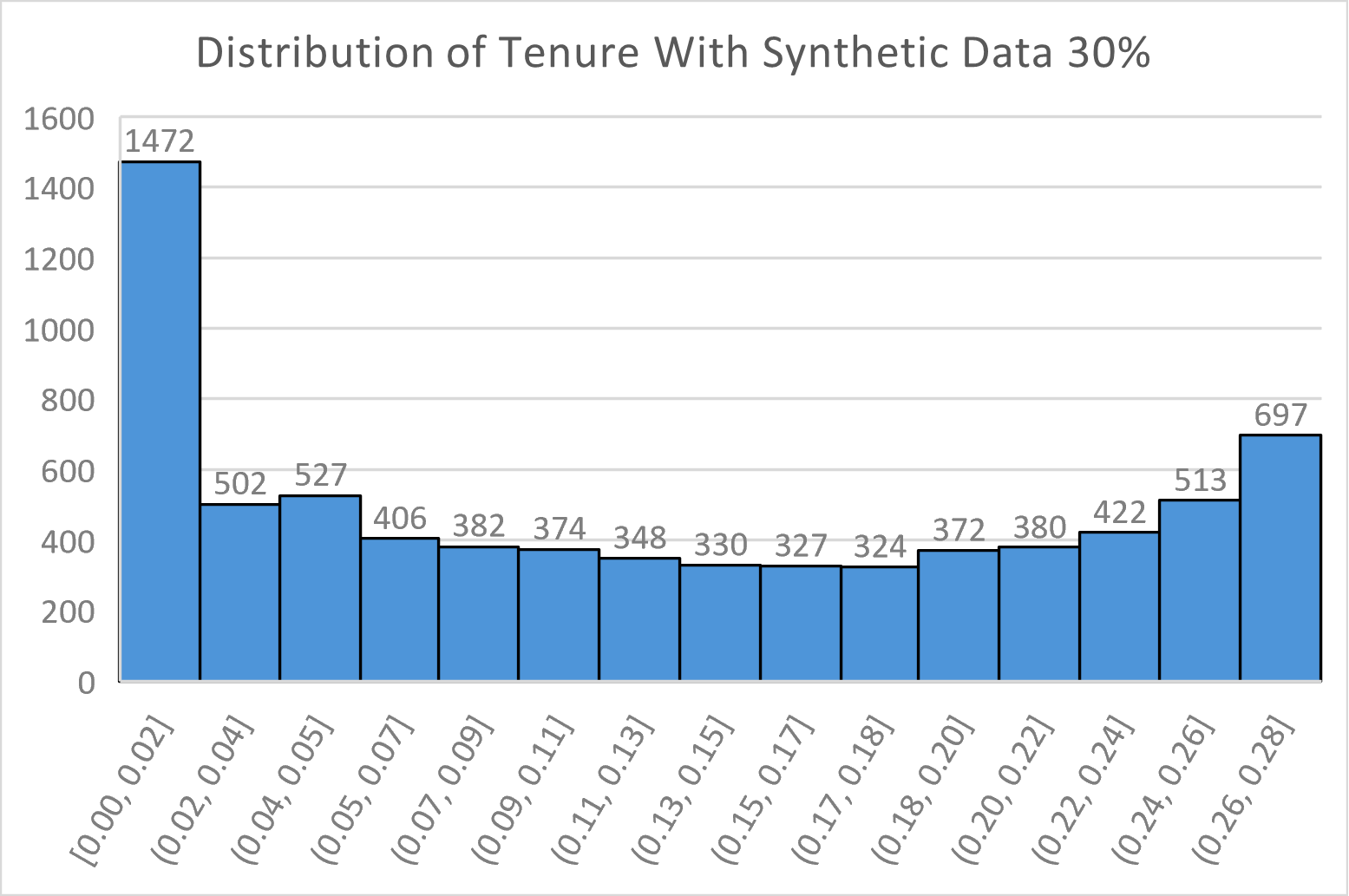} 
\caption{}
\label{qsmote_Fig7a}
\end{subfigure}\hfill
\begin{subfigure}{0.33\linewidth}
\includegraphics[width=\linewidth]{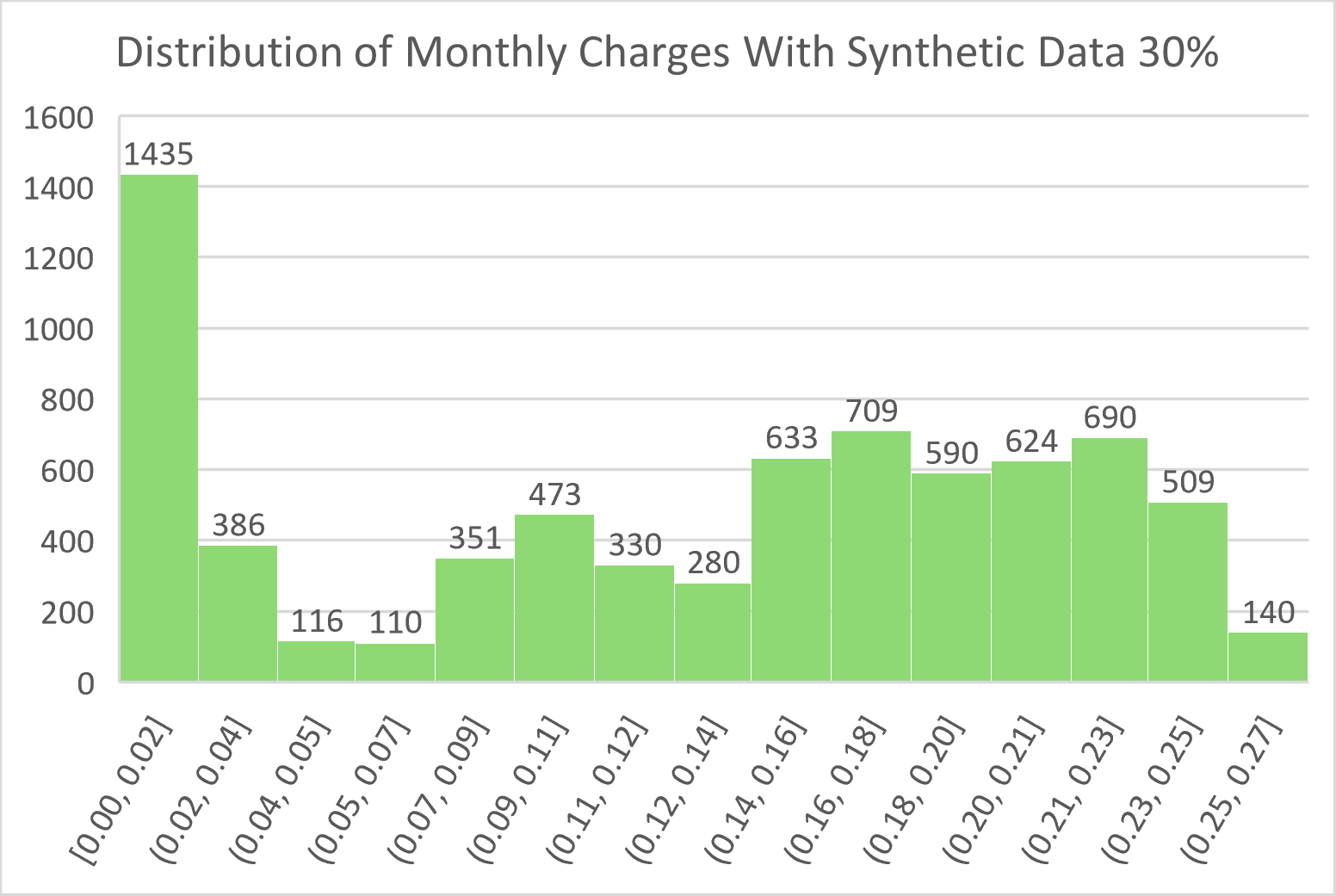} 
\caption{}
\label{qsmote_Fig7b}
\end{subfigure}\hfill
\begin{subfigure}{0.33\linewidth}
\includegraphics[width=\linewidth]{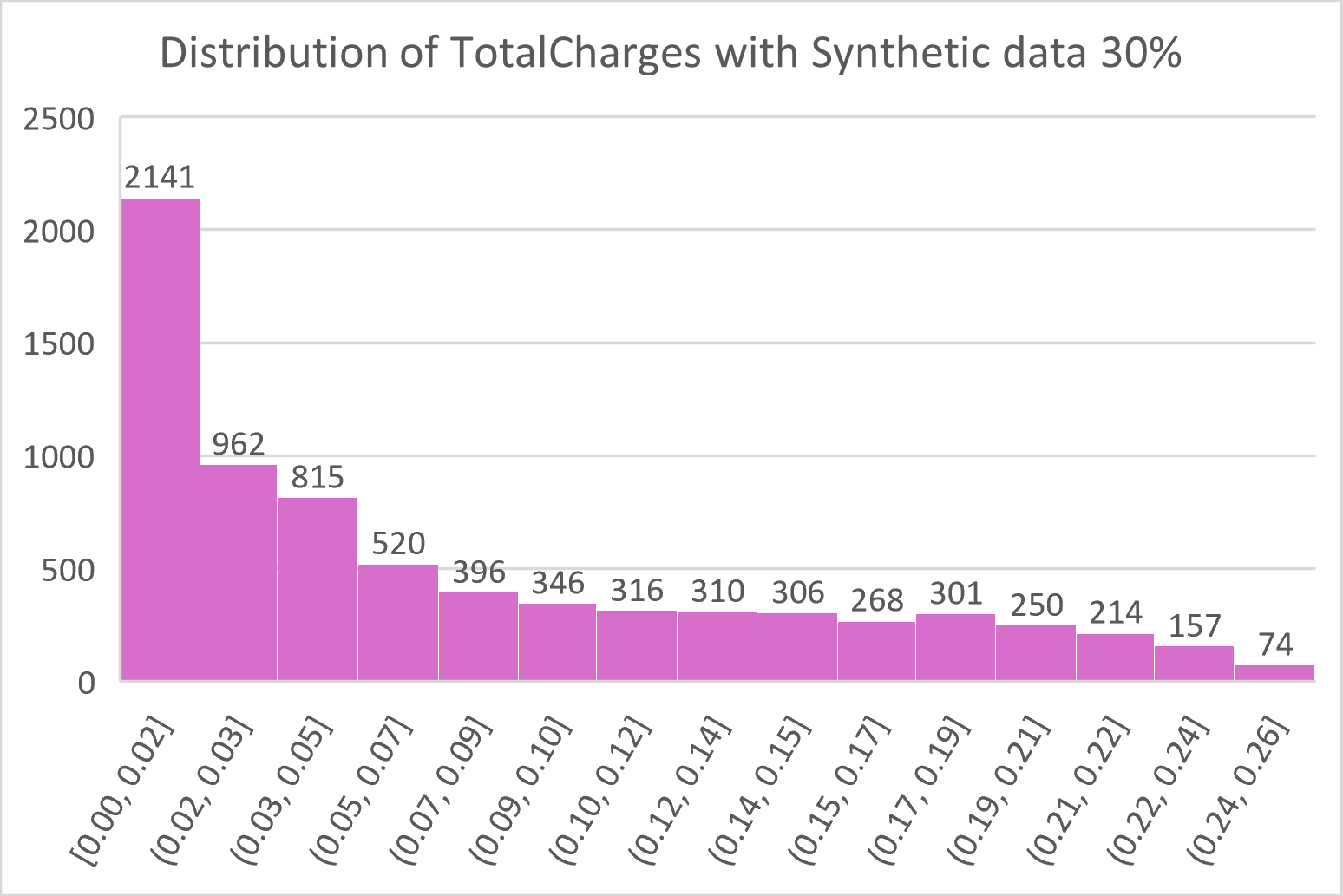} 
\caption{}
\label{qdctc_Fig7c}
\end{subfigure}\hfill
\caption{Plot illustrating distribution of 3 columns with induction of synthetic datapoints with overall $30\%$ minority : (a) Tenure, (b) MonthlyCharges, and (c) TotalCharges.}
\label{qsmote_Fig8}
\end{figure}

\begin{figure}[H]
\centering
\begin{subfigure}{0.33\linewidth}
\includegraphics[width=\linewidth]{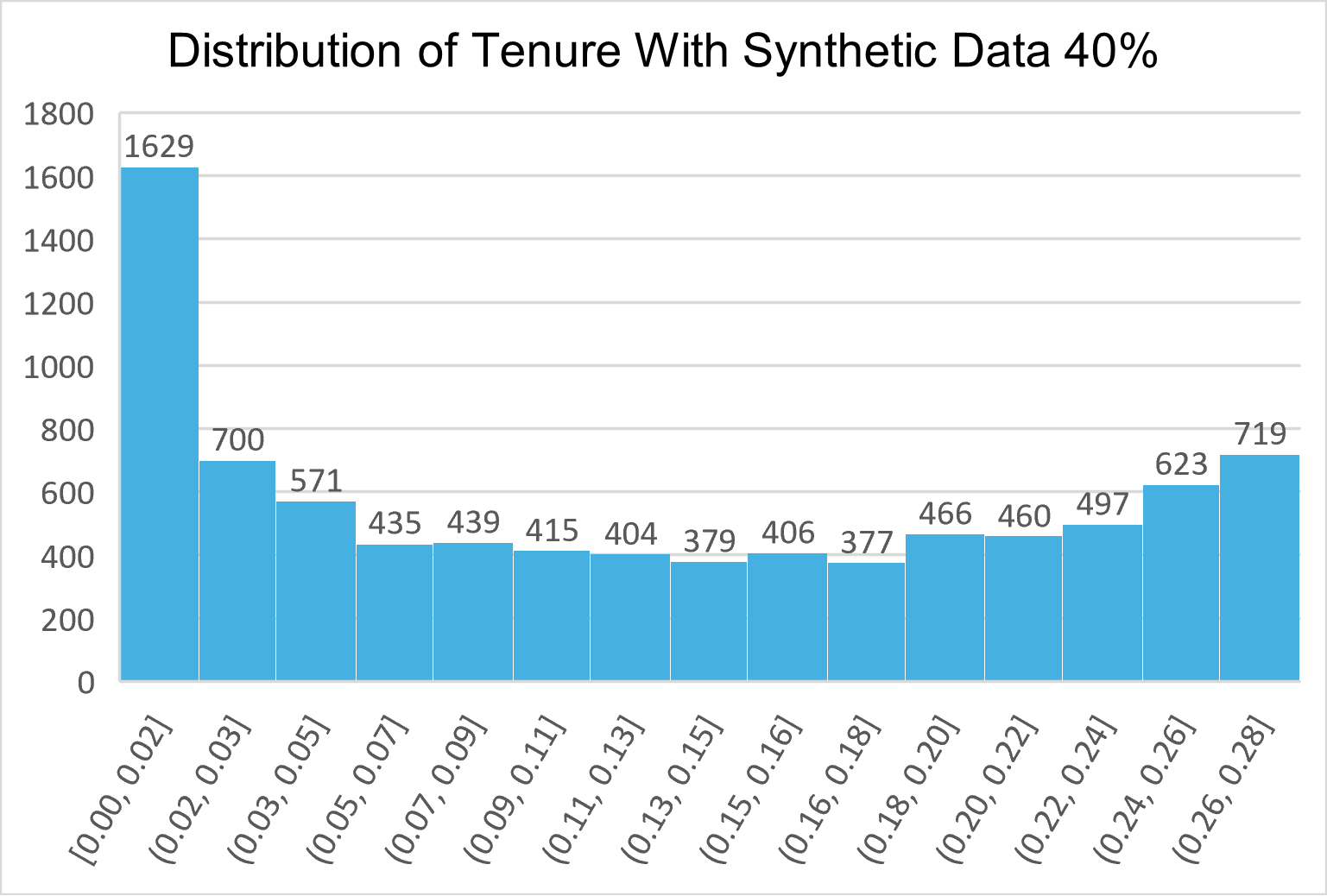} 
\caption{}
\label{qsmote_Fig8a}
\end{subfigure}\hfill
\begin{subfigure}{0.33\linewidth}
\includegraphics[width=\linewidth]{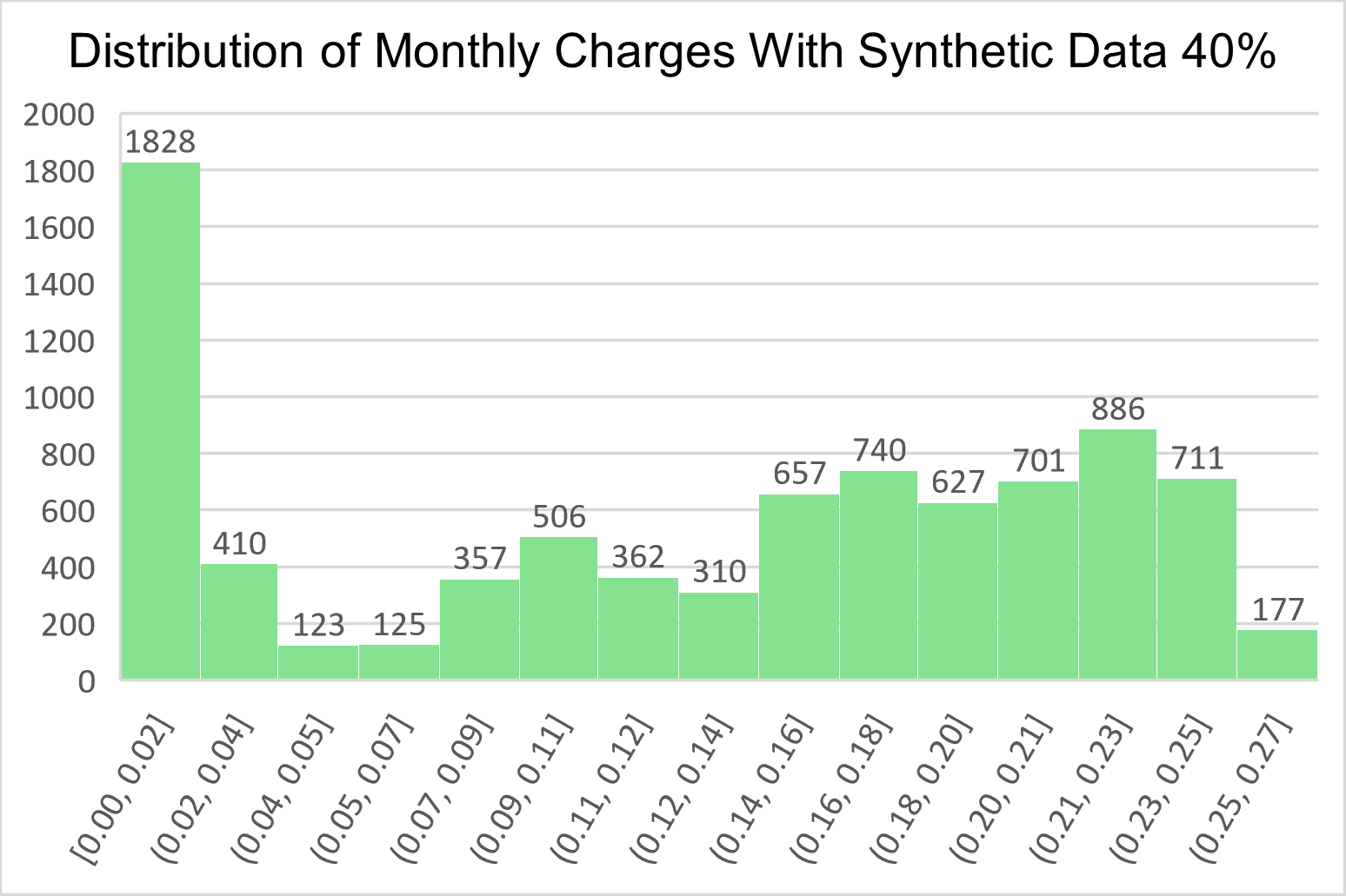} 
\caption{}
\label{qsmote_Fig8b}
\end{subfigure}\hfill
\begin{subfigure}{0.33\linewidth}
\includegraphics[width=\linewidth]{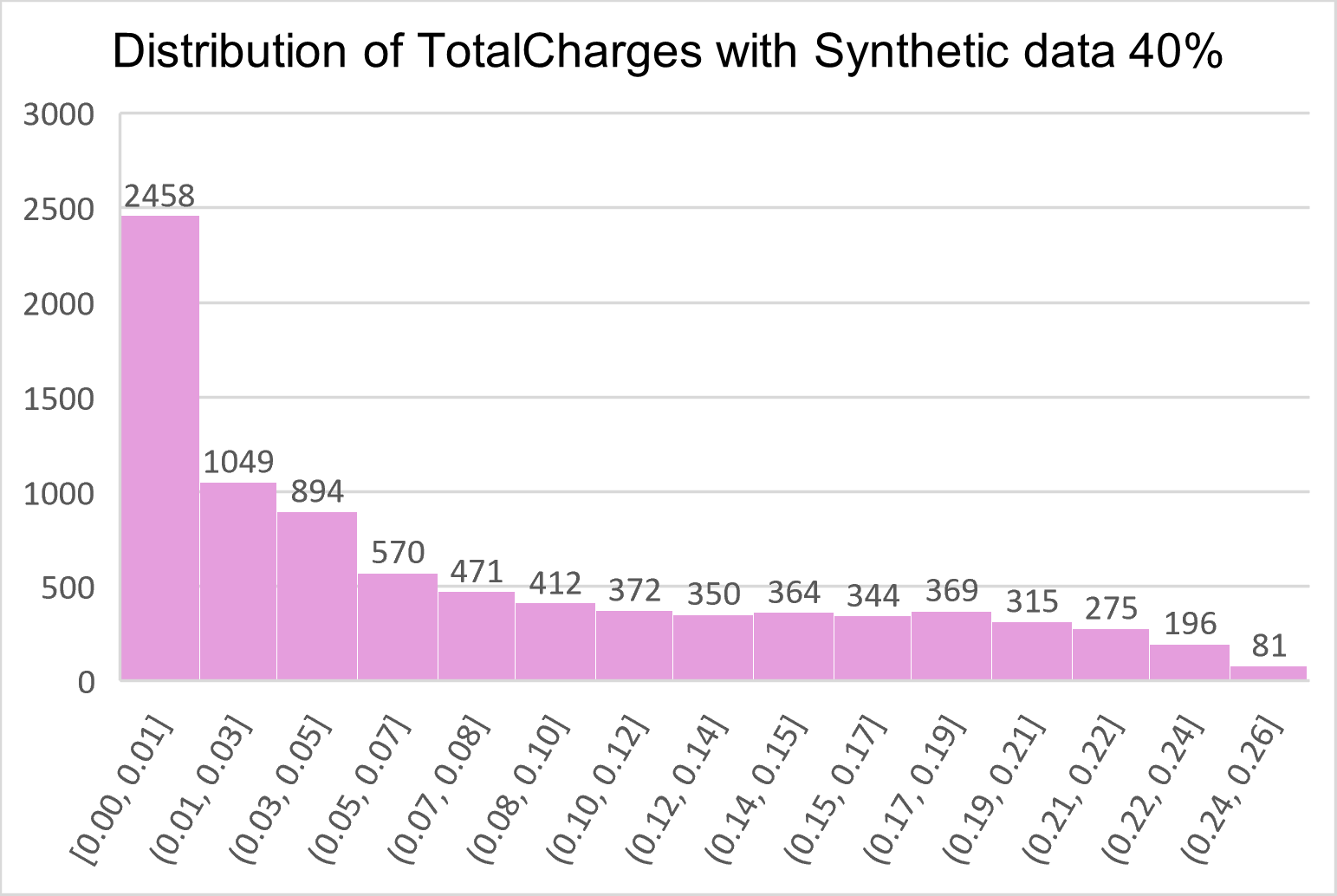} 
\caption{}
\label{qdctc_Fig8c}
\end{subfigure}\hfill
\caption{Plot illustrating distribution of 3 columns with induction of synthetic data points with overall $40\%$ minority: (a) Tenure, (b) MonthlyCharges, and (c) TotalCharges.}
\label{qsmote_Fig9}
\end{figure}

\begin{figure}[H]
\centering
\begin{subfigure}{0.33\linewidth}
\includegraphics[width=\linewidth]{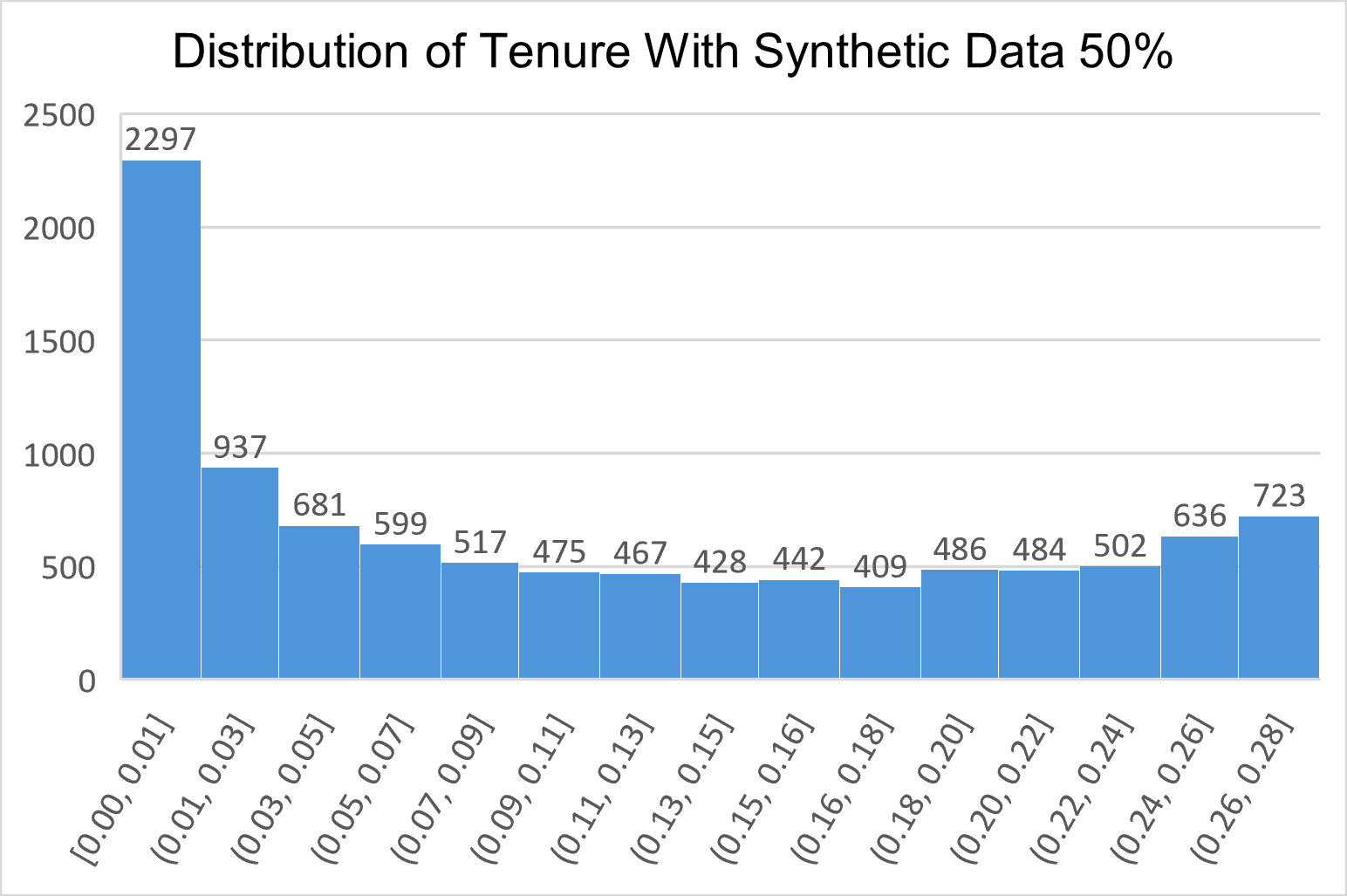} 
\caption{}
\label{qsmote_Fig9a}
\end{subfigure}\hfill
\begin{subfigure}{0.33\linewidth}
\includegraphics[width=\linewidth]{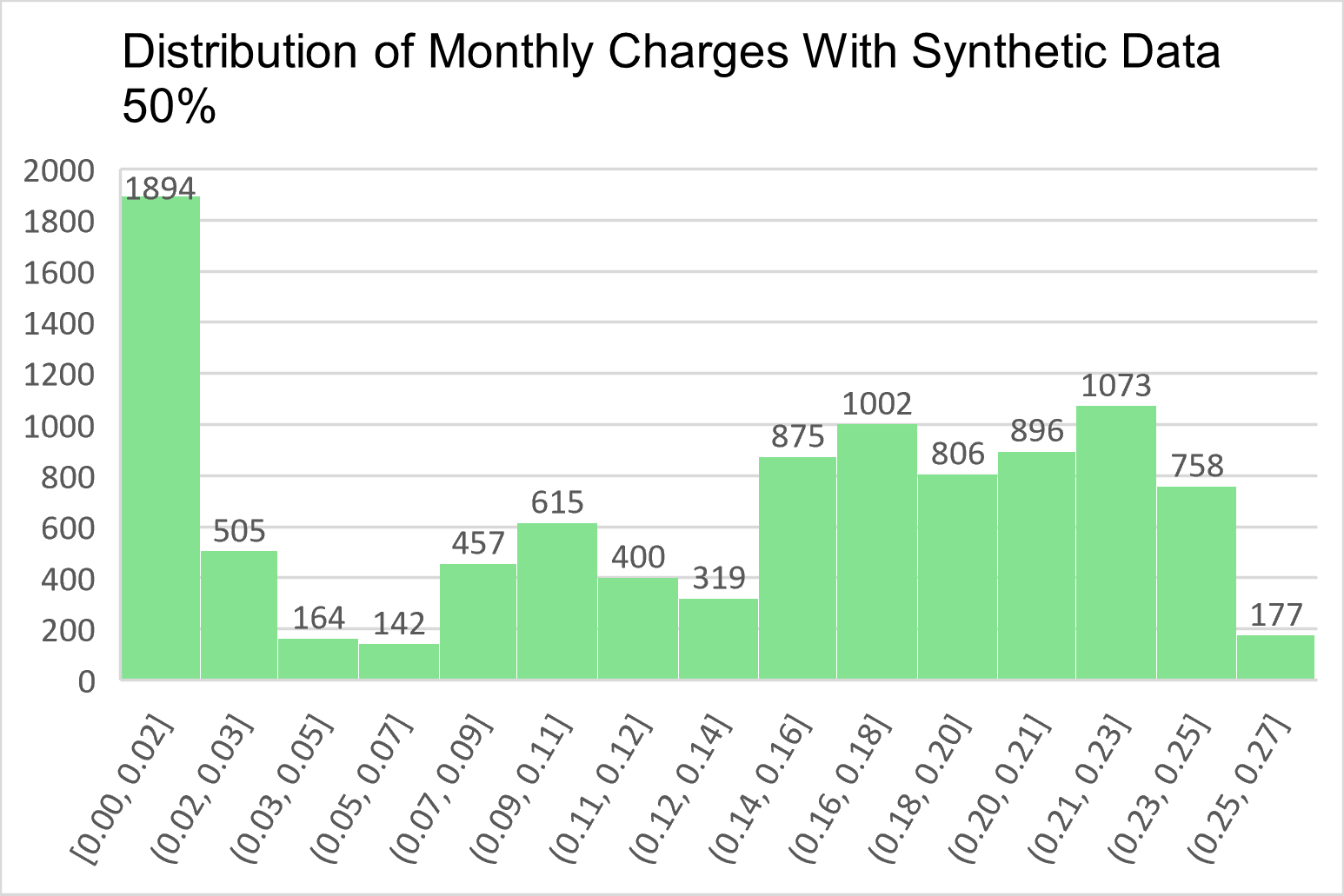} 
\caption{}
\label{qsmote_Fig9b}
\end{subfigure}\hfill
\begin{subfigure}{0.33\linewidth}
\includegraphics[width=\linewidth]{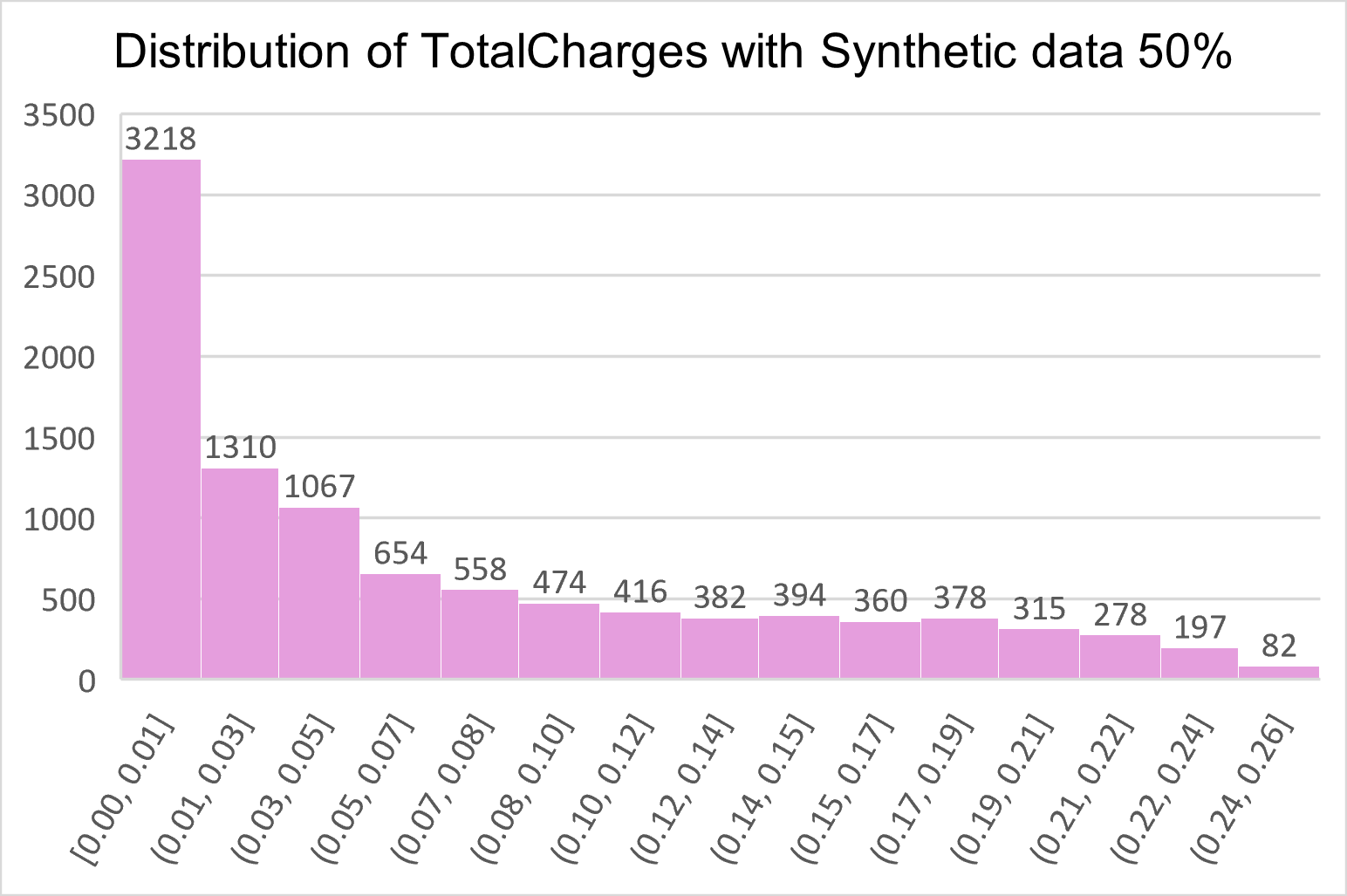} 
\caption{}
\label{qdctc_Fig9c}
\end{subfigure}\hfill
\caption{Plot illustrating distribution of 3 columns with induction of synthetic data points with overall $50\%$ minority : (a) Tenure, (b) MonthlyCharges, and (c) TotalCharges.}
\label{qsmote_Fig10}
\end{figure}

\subsubsection{Observation from generated data}

Following the generation of synthetic data points by rotation, our next step is to examine the general distribution of important variables throughout the whole population. The objective is to assess if the introduction of artificial data points has caused any significant statistical deviation in the distribution of the variable. The figures \ref{qsmote_Fig7}, \ref{qsmote_Fig8}, \ref{qsmote_Fig9}, and \ref{qsmote_Fig10} illustrate the distribution of three important variables in the dataset: Tenure, MonthlyCharges, and Total charges. The distribution before the induction of synthetic data points is shown in Fig. \ref{qsmote_Fig7}. The distributions following the induction of synthetic data points, resulting in total minority percentages of $30$, $40$, and $50$, are shown in figures \ref{qsmote_Fig8}, \ref{qsmote_Fig9}, and \ref{qsmote_Fig10} accordingly. After applying SMOTE, we can confidently state that there is a little distortion to the distribution of variables, but the bins have increased in size. The use of relatively modest angles during rotation prevents any significant deformation to the geometry of the distribution. By comparing the charts depicting the variables after using the SMOTE technique, we see a progressive rise in the values within each category, ranging from $30\%$ to $50\%$. This confirms the successful use of the SMOTE method.

\subsubsection{Applying Classification Models}
In order to comprehensively evaluate the effectiveness of the Synthetic Minority Over-sampling Technique (SMOTE) in addressing class imbalances, our research used two classification models, namely Random Forest and Logistic Regression, on the Telecom Churn Dataset. The selection of these models was made to assess the influence of using SMOTE on the performance of the models, particularly in situations characterized by an imbalance in class distribution.
The Random Forest algorithm is well recognized for its ability to efficiently handle skewed datasets. This model utilizes ensemble learning by creating multiple decision trees and aggregating their predictions to mitigate overfitting. The algorithm natively addresses class imbalances by using techniques such as bootstrap sampling and adjusting its class weights parameter to enhance sensitivity towards the minority class. This eliminates the requirement for external interventions like SMOTE \cite{breiman_random_2001}.
On the other hand, Logistic Regression, a model well regarded for its simplicity and effectiveness in situations where binary classification is needed, was selected to provide a contrasting analytical viewpoint. The classification strategy of Logistic Regression, which entails estimating the likelihood that a certain data point belongs to a specific class, does not inherently tackle the issue of class imbalance \cite{Hosmer_2013_logistic}. This attribute makes it a perfect contender for evaluating the immediate impacts of SMOTE on model efficacy, providing valuable observations on how SMOTE might augment a model's capacity to identify the underrepresented class in unbalanced datasets.

The research seeks to evaluate the efficiency of the SMOTE method across various modeling techniques by comparing the performances of these models before and after their deployment. An investigation of SMOTE's adaptability in enhancing classification results is crucial, especially for models such as Logistic Regression that lack inherent methods for addressing data imbalances \cite{chawla_smote_2002}
.

To evaluate the model, we have used the Confusion Matrix, Accuracy, Precision, Recall, F1-Score, and the Area Under the Receiver Operating Characteristic Curve (AUC-ROC). Below are the model evaluation charts for the Random Forest Model followed by the Logistic Regression Model.

\begin{figure}[H]
\centering
\begin{subfigure}{0.5\linewidth}
\includegraphics[width=\linewidth]{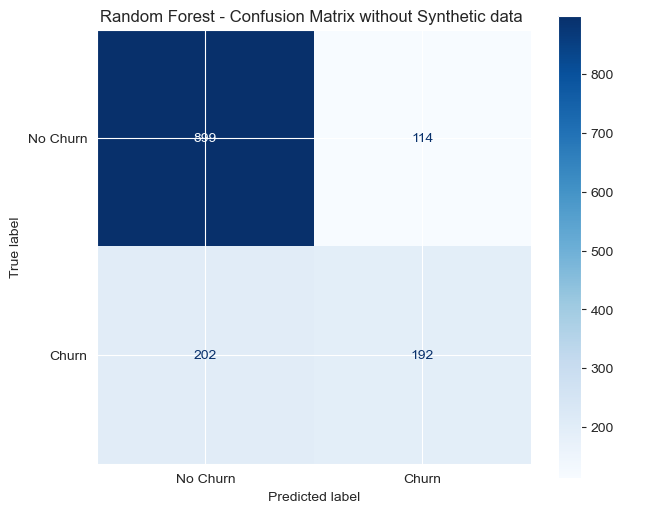} 
\caption{}
\label{qsmote_Fig10a}
\end{subfigure}\hfill
\begin{subfigure}{0.5\linewidth}
\includegraphics[width=\linewidth]{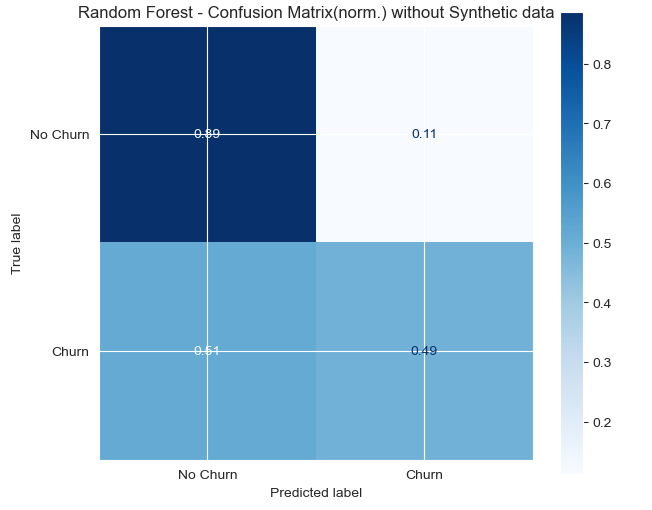} 
\caption{}
\label{qsmote_Fig10b}
\end{subfigure}\hfill
\begin{subfigure}{0.5\linewidth}
\includegraphics[width=\linewidth]{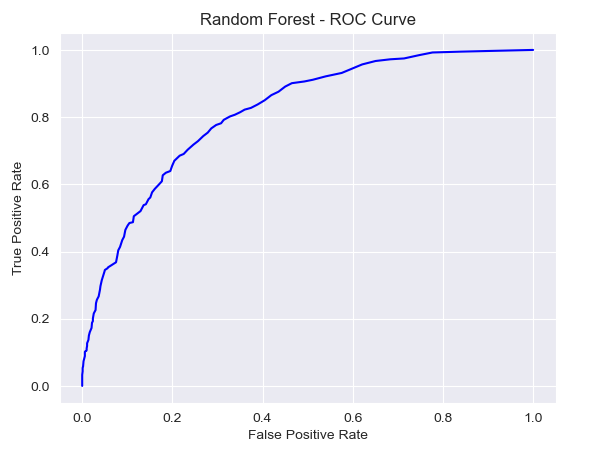} 
\caption{}
\label{qsmote_Fig10c}
\end{subfigure}\hfill
\begin{subfigure}{0.5\linewidth}
\includegraphics[width=\linewidth]{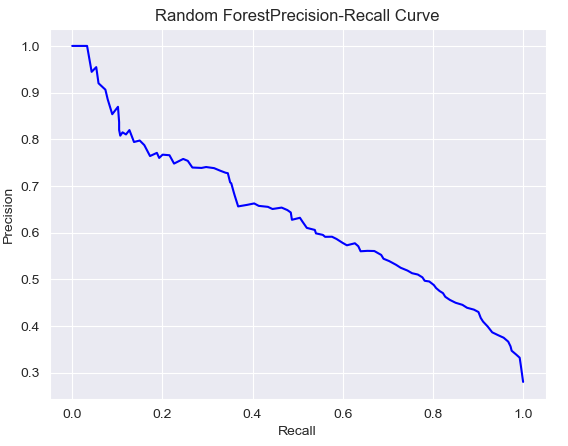} 
\caption{}
\label{qsmote_Fig10d}
\end{subfigure}\hfill
\caption{Plot illustrating Model Charts for random forest model with out SMOTE. (a) Confusion Matrix Random Forest Model, (b) Normalised Confusion Matrix Random Forest Model, (c) AUC-ROC Random Foerest Model, (d) Precision Recall Curve Random Forest Model.}
\label{qsmote_Fig11}
\end{figure}

\begin{figure}[H]
\centering
\begin{subfigure}{0.5\linewidth}
\includegraphics[width=\linewidth]{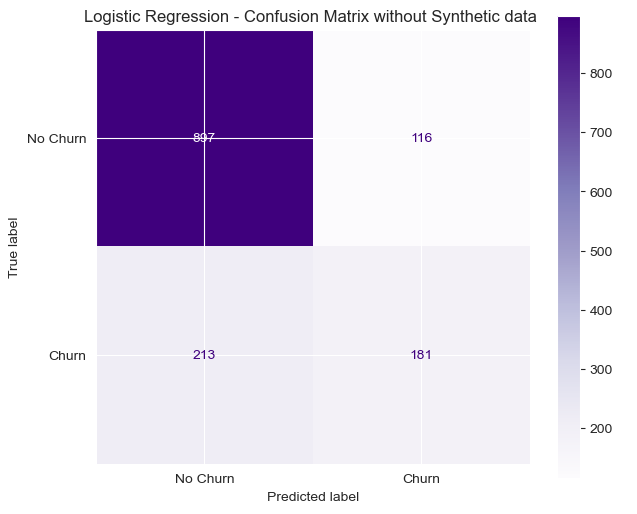} 
\caption{}
\label{qsmote_Fig11a}
\end{subfigure}\hfill
\begin{subfigure}{0.5\linewidth}
\includegraphics[width=\linewidth]{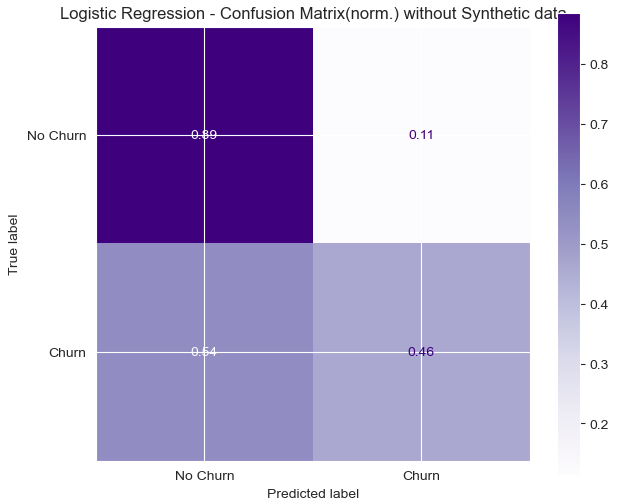} 
\caption{}
\label{qsmote_Fig11b}
\end{subfigure}\hfill
\begin{subfigure}{0.5\linewidth}
\includegraphics[width=\linewidth]{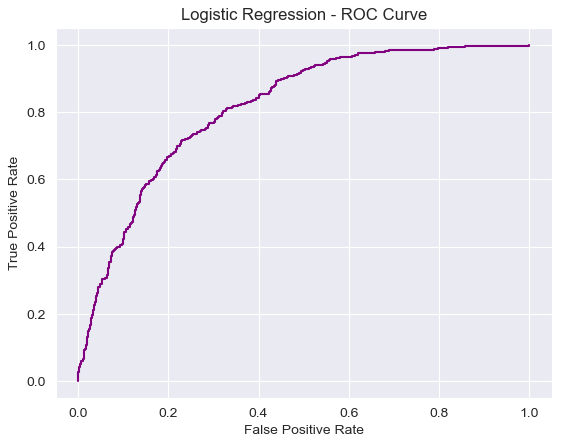} 
\caption{}
\label{qsmote_Fig11c}
\end{subfigure}\hfill
\begin{subfigure}{0.5\linewidth}
\includegraphics[width=\linewidth]{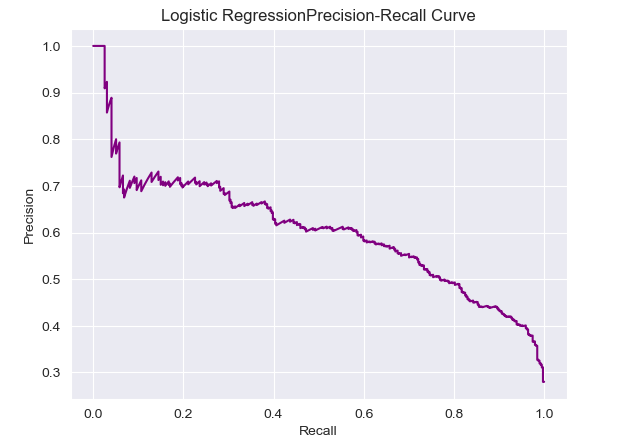} 
\caption{}
\label{qsmote_Fig11d}
\end{subfigure}\hfill
\caption{Plot illustrating Model Charts for logistic regression model without SMOTE. (a) Confusion Matrix Logistic Regression Model, (b) Normalised Confusion Matrix Logistic Regression Model, (c) AUC-ROC Logistic Regression Model, (d) Precision Recall Curve Logistic Regression Model.}
\label{qsmote_Fig12}
\end{figure}
\begin{figure}[H]
\centering
\begin{subfigure}{0.5\linewidth}
\includegraphics[width=\linewidth]{Figures/Confusion_Matrix_RF.png} 
\caption{}
\label{qsmote_Fig12a}
\end{subfigure}\hfill
\begin{subfigure}{0.5\linewidth}
\includegraphics[width=\linewidth]{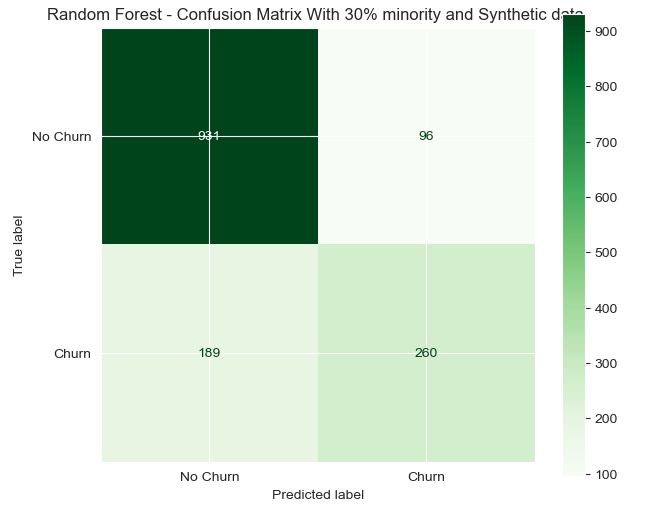} 
\caption{}
\label{qsmote_Fig12b}
\end{subfigure}\hfill
\begin{subfigure}{0.5\linewidth}
\includegraphics[width=\linewidth]{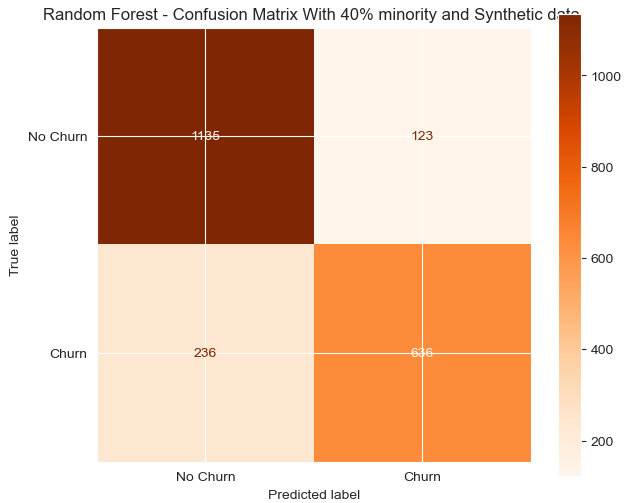} 
\caption{}
\label{qsmote_Fig12c}
\end{subfigure}\hfill
\begin{subfigure}{0.5\linewidth}
\includegraphics[width=\linewidth]{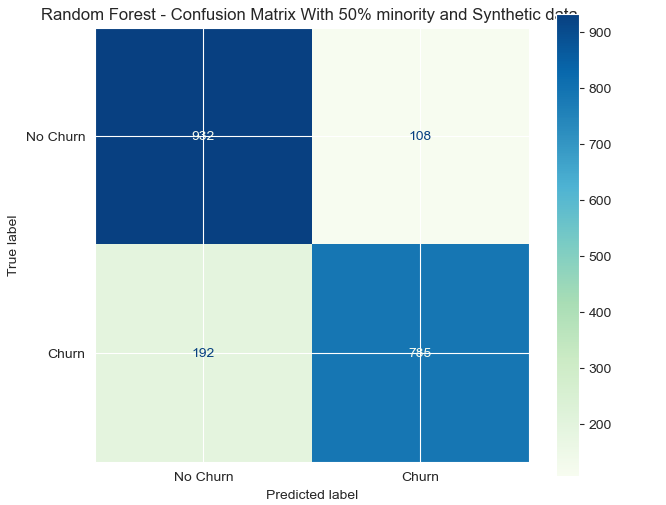} 
\caption{}
\label{qsmote_Fig12d}
\end{subfigure}\hfill
\caption{Plot illustrating Confusion Matrix for random forest model with and without smote for comparison. (a) Confusion Matrix Random Forest Model without smote, (b) Confusion Matrix Random Forest Model with smote and $30\%$ Minority, (c) Confusion Matrix Random Forest Model with smote and $40\%$ Minority, (d) Confusion Matrix Random Forest Model with smote and $50\%$ Minority.}
\label{qsmote_Fig13}
\end{figure}

\begin{figure}[H]
\centering
\begin{subfigure}{0.5\linewidth}
\includegraphics[width=\linewidth]{Figures/Confusion_Matrix_RF_Normalised.png} 
\caption{}
\label{qsmote_Fig13a}
\end{subfigure}\hfill
\begin{subfigure}{0.5\linewidth}
\includegraphics[width=\linewidth]{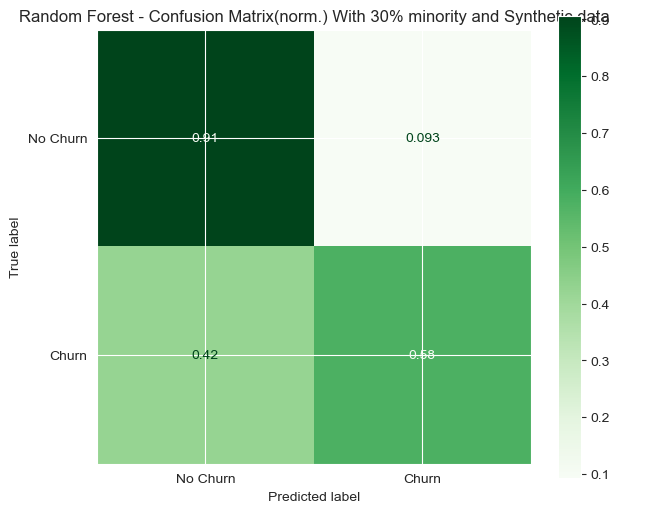} 
\caption{}
\label{qsmote_Fig13b}
\end{subfigure}\hfill
\begin{subfigure}{0.5\linewidth}
\includegraphics[width=\linewidth]{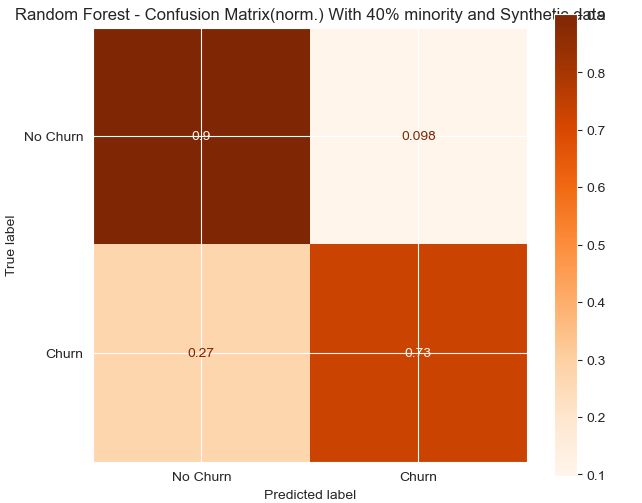} 
\caption{}
\label{qsmote_Fig13c}
\end{subfigure}\hfill
\begin{subfigure}{0.5\linewidth}
\includegraphics[width=\linewidth]{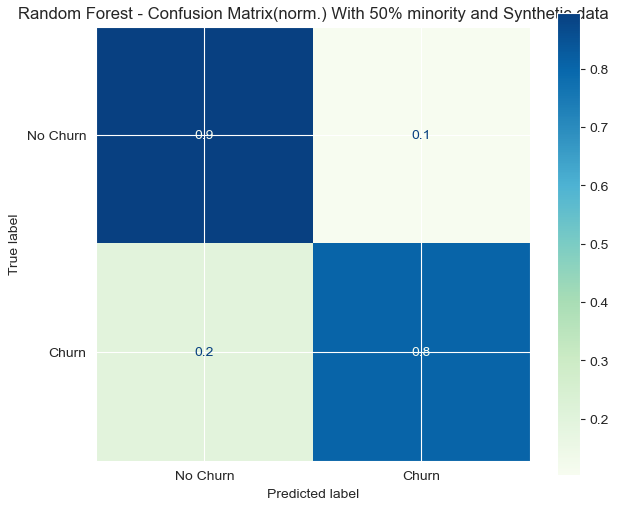} 
\caption{}
\label{qsmote_Fig13d}
\end{subfigure}\hfill
\caption{Plot illustrating Normalized Confusion Matrix for random forest model with and without smote for comparison. (a) Normalized Confusion Matrix Random Forest Model without smote, (b) Normalized Confusion Matrix Random Forest Model with smote and 30\% Minority, (c) Normalized Confusion Matrix Random Forest Model with smote and 40\% Minority, (d) Normalized Confusion Matrix Random Forest Model with smote and 50\% Minority.}
\label{qsmote_Fig14}
\end{figure}

\begin{figure}[H]
\centering
\begin{subfigure}{0.5\linewidth}
\includegraphics[width=\linewidth]{Figures/ROC_AUC_RF.png} 
\caption{}
\label{qsmote_Fig14a}
\end{subfigure}\hfill
\begin{subfigure}{0.5\linewidth}
\includegraphics[width=\linewidth]{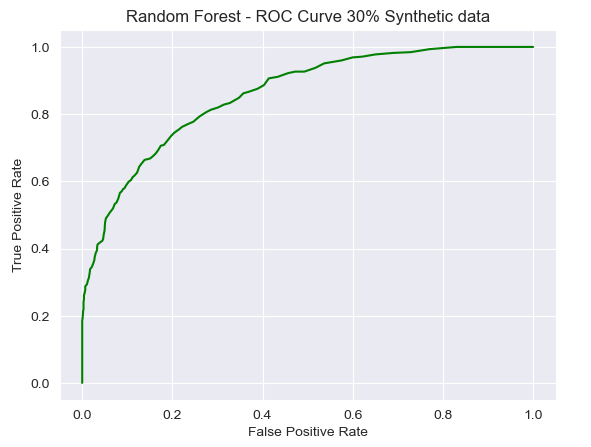} 
\caption{}
\label{qsmote_Fig14b}
\end{subfigure}\hfill
\begin{subfigure}{0.5\linewidth}
\includegraphics[width=\linewidth]{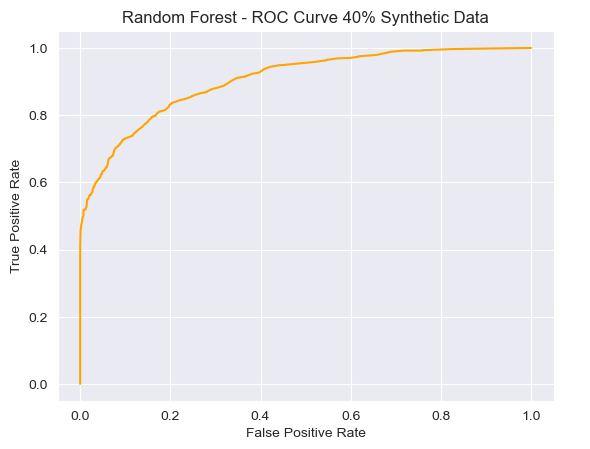} 
\caption{}
\label{qsmote_Fig14c}
\end{subfigure}\hfill
\begin{subfigure}{0.5\linewidth}
\includegraphics[width=\linewidth]{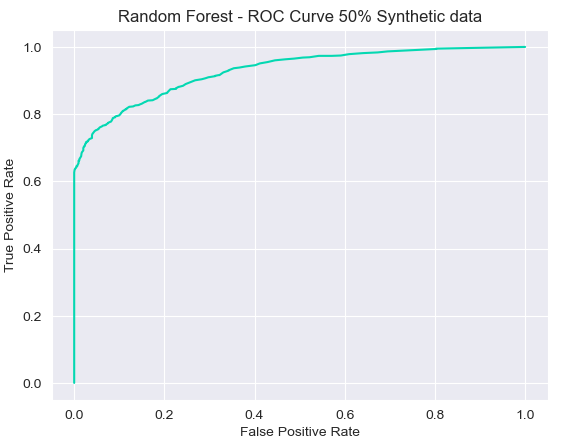} 
\caption{}
\label{qsmote_Fig14d}
\end{subfigure}\hfill
\caption{Plot illustrating Area Under Receiver Operating Characteristic Curve (AUC-ROC) for random forest model with and without smote for comparison. (a) AUC-ROC Random Forest Model without smote, (b) AUC-ROC Random Forest Model with smote and 30\% Minority, (c) AUC-ROC Random Forest Model with smote and 40\% Minority, (d) AUC-ROC Random Forest Model with smote and 50\% Minority.}
\label{qsmote_Fig15}
\end{figure}

\begin{figure}[H]
\centering
\begin{subfigure}{0.5\linewidth}
\includegraphics[width=\linewidth]{Figures/Presicion_Recall_RF.png} 
\caption{}
\label{qsmote_Fig15a}
\end{subfigure}\hfill
\begin{subfigure}{0.5\linewidth}
\includegraphics[width=\linewidth]{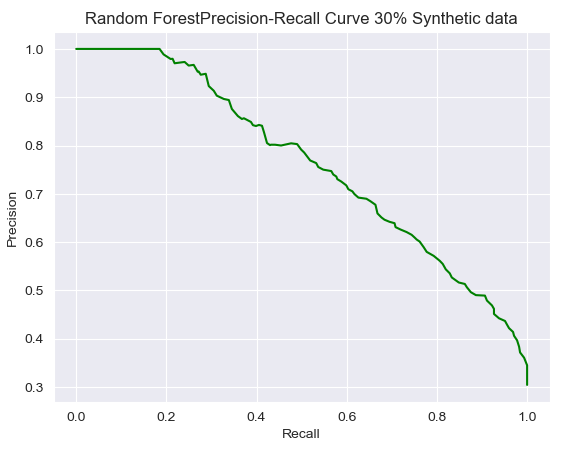} 
\caption{}
\label{qsmote_Fig15b}
\end{subfigure}\hfill
\begin{subfigure}{0.5\linewidth}
\includegraphics[width=\linewidth]{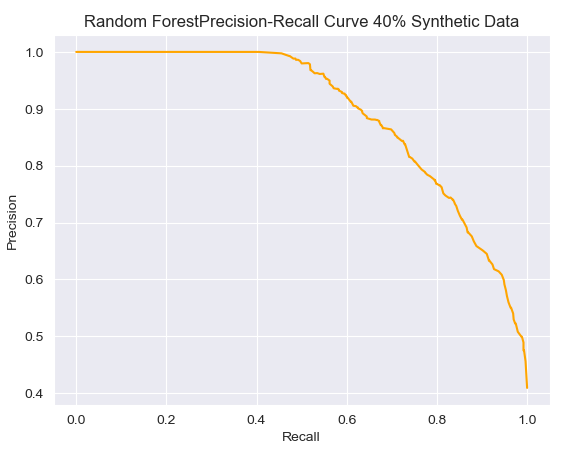} 
\caption{}
\label{qsmote_Fig15c}
\end{subfigure}\hfill
\begin{subfigure}{0.5\linewidth}
\includegraphics[width=\linewidth]{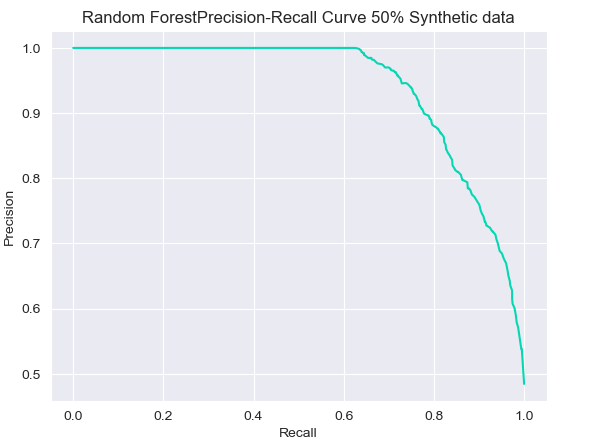} 
\caption{}
\label{qsmote_Fig15d}
\end{subfigure}\hfill
\caption{Plot illustrating Precision-Recall Curve (AUC) for random forest model with and without smote for comparison. (a) Precision-Recall Curve (AUC) Random Forest Model without smote, (b) Precision-Recall Curve (AUC) Random Forest Model with smote and 30\% Minority, (c) Precision-Recall Curve (AUC) Random Forest Model with smote and 40\% Minority, (d) Precision-Recall Curve (AUC) Random Forest Model with smote and 50\% Minority.}
\label{qsmote_Fig16}
\end{figure}
\begin{figure}[H]
\centering
\begin{subfigure}{0.5\linewidth}
\includegraphics[width=\linewidth]{Figures/Confusion_Matrix_LR.png} 
\caption{}
\label{qsmote_Fig16a}
\end{subfigure}\hfill
\begin{subfigure}{0.5\linewidth}
\includegraphics[width=\linewidth]{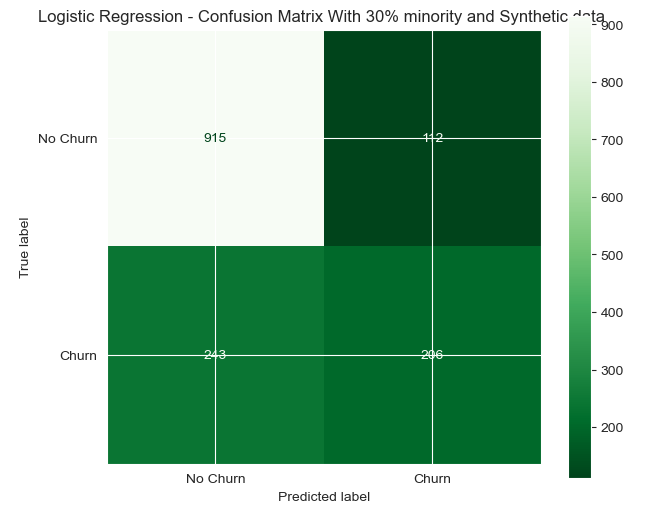} 
\caption{}
\label{qsmote_Fig16b}
\end{subfigure}\hfill
\begin{subfigure}{0.5\linewidth}
\includegraphics[width=\linewidth]{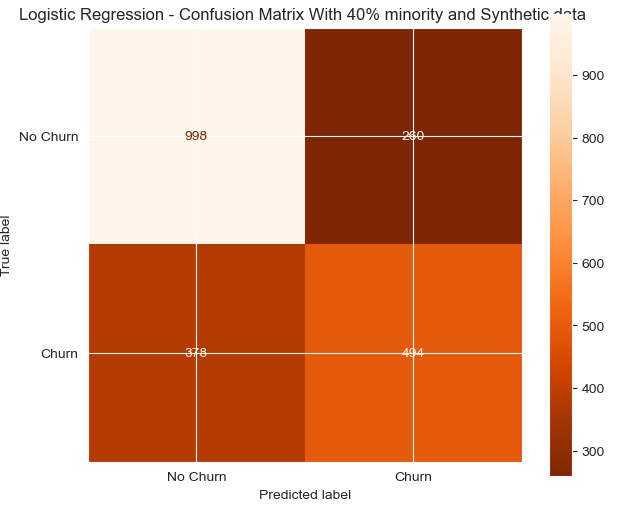} 
\caption{}
\label{qsmote_Fig16c}
\end{subfigure}\hfill
\begin{subfigure}{0.5\linewidth}
\includegraphics[width=\linewidth]{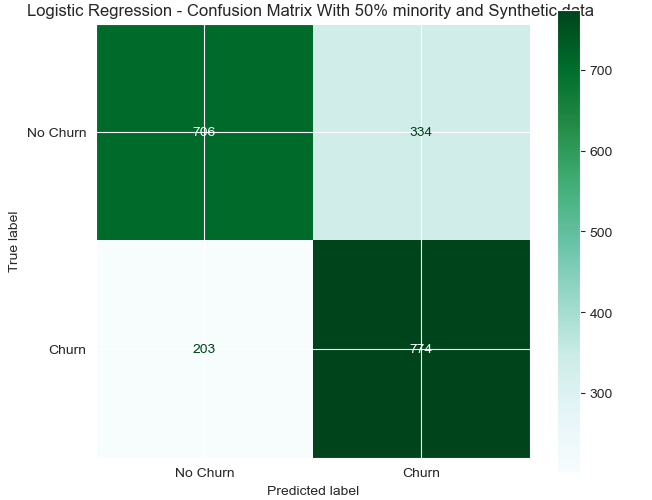} 
\caption{}
\label{qsmote_Fig16d}
\end{subfigure}\hfill
\caption{Plot illustrating Confusion Matrix for Logistic Regression model with and without smote for comparison. (a) Confusion Matrix Logistic Regression Model without smote, (b) Confusion Matrix Logistic Regression Model with smote and 30\% Minority, (c) Confusion Matrix Logistic Regression Model with smote and 40\% Minority, (d) Confusion Matrix Logistic Regression Model with smote and 50\% Minority.}
\label{qsmote_Fig17}
\end{figure}

\begin{figure}[H]
\centering
\begin{subfigure}{0.5\linewidth}
\includegraphics[width=\linewidth]{Figures/Confusion_Matrix_LR_Normalised.png} 
\caption{}
\label{qsmote_Fig17a}
\end{subfigure}\hfill
\begin{subfigure}{0.5\linewidth}
\includegraphics[width=\linewidth]{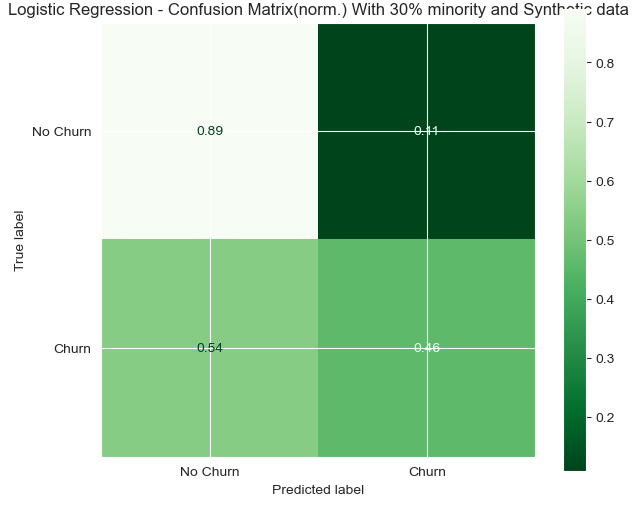} 
\caption{}
\label{qsmote_Fig17b}
\end{subfigure}\hfill
\begin{subfigure}{0.5\linewidth}
\includegraphics[width=\linewidth]{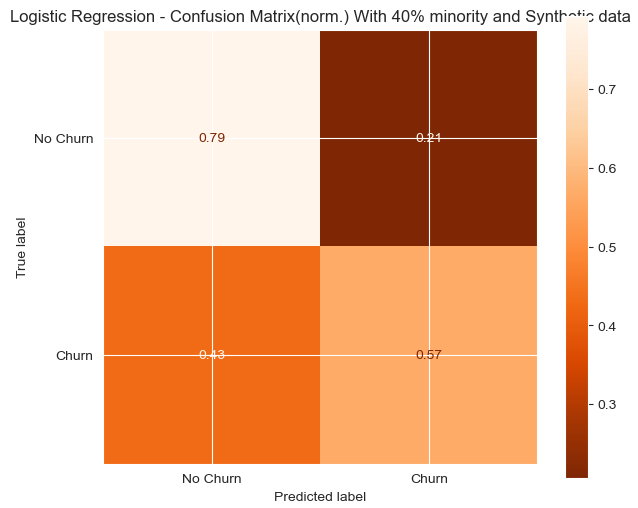} 
\caption{}
\label{qsmote_Fig17c}
\end{subfigure}\hfill
\begin{subfigure}{0.5\linewidth}
\includegraphics[width=\linewidth]{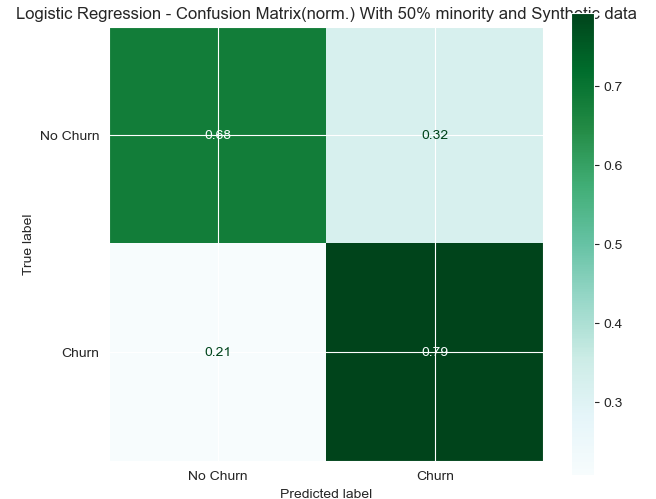} 
\caption{}
\label{qsmote_Fig17d}
\end{subfigure}\hfill
\caption{Plot illustrating Normalized Confusion Matrix for Logistic Regression model with and without smote for comparison. (a) Normalized Confusion Matrix Logistic Regression Model without smote, (b) Normalized Confusion Matrix Logistic Regression Model with smote and 30\% Minority, (c) Normalized Confusion Matrix Logistic Regression Model with smote and 40\% Minority, (d) Normalized Confusion Matrix Logistic Regression Model with smote and 50\% Minority.}
\label{qsmote_Fig18}
\end{figure}

\begin{figure}[H]
\centering
\begin{subfigure}{0.5\linewidth}
\includegraphics[width=\linewidth]{Figures/ROC_AUC_LR.png} 
\caption{}
\label{qsmote_Fig18a}
\end{subfigure}\hfill
\begin{subfigure}{0.5\linewidth}
\includegraphics[width=\linewidth]{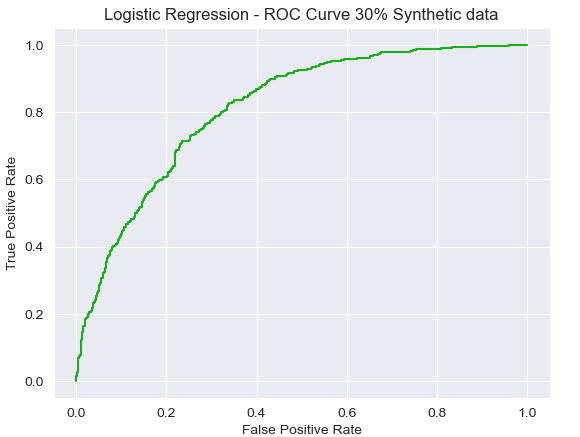} 
\caption{}
\label{qsmote_Fig18b}
\end{subfigure}\hfill
\begin{subfigure}{0.5\linewidth}
\includegraphics[width=\linewidth]{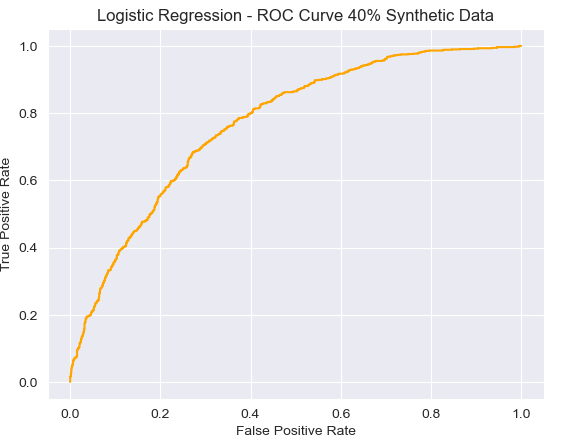} 
\caption{}
\label{qsmote_Fig18c}
\end{subfigure}\hfill
\begin{subfigure}{0.5\linewidth}
\includegraphics[width=\linewidth]{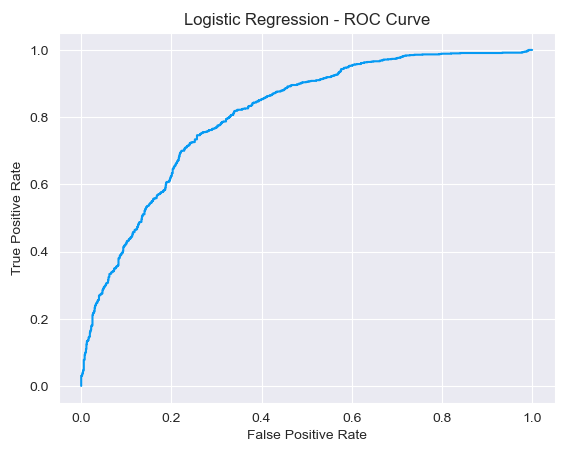} 
\caption{}
\label{qsmote_Fig18d}
\end{subfigure}\hfill
\caption{Plot illustrating Area Under Receiver Operating Characteristic Curve (AUC-ROC) for Logistic Regression model with and without smote for comparison. (a) AUC-ROC Logistic Regression Model without smote, (b) AUC-ROC Logistic Regression Model with smote and 30\% Minority, (c) AUC-ROC Logistic Regression Model with smote and 40\% Minority, (d) AUC-ROC Logistic Regression Model with smote and 50\% Minority.}
\label{qsmote_Fig19}
\end{figure}

\begin{figure}[H]
\centering
\begin{subfigure}{0.5\linewidth}
\includegraphics[width=\linewidth]{Figures/Presicion_Recall_LR.png} 
\caption{}
\label{qsmote_Fig19a}
\end{subfigure}\hfill
\begin{subfigure}{0.5\linewidth}
\includegraphics[width=\linewidth]{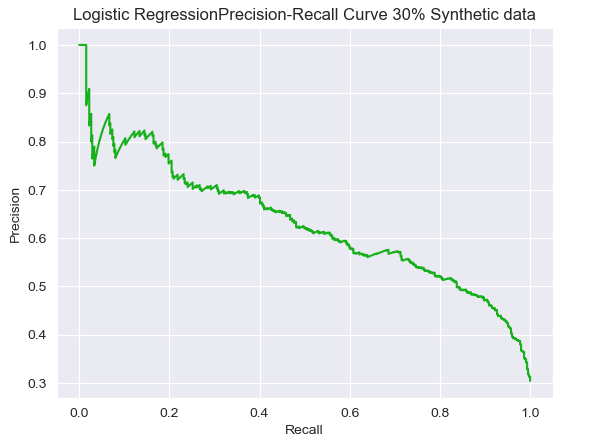} 
\caption{}
\label{qsmote_Fig19b}
\end{subfigure}\hfill
\begin{subfigure}{0.5\linewidth}
\includegraphics[width=\linewidth]{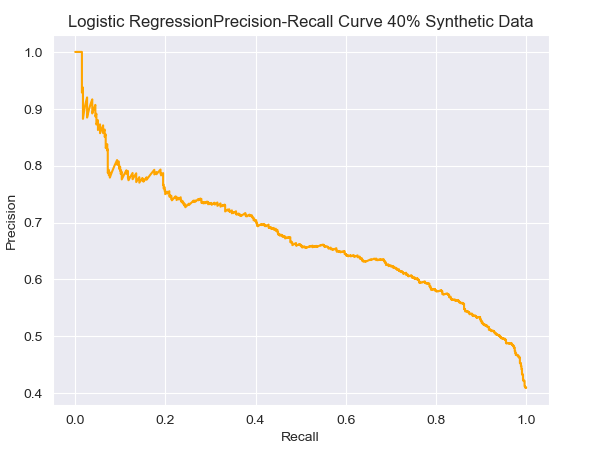} 
\caption{}
\label{qsmote_Fig19c}
\end{subfigure}\hfill
\begin{subfigure}{0.5\linewidth}
\includegraphics[width=\linewidth]{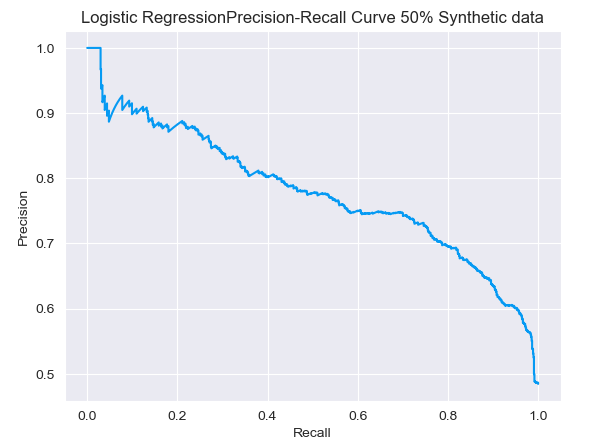} 
\caption{}
\label{qsmote_Fig19d}
\end{subfigure}\hfill
\caption{Plot illustrating Precision-Recall Curve (AUC) for Logistic Regression model with and without smote for comparison. (a) Precision-Recall Curve (AUC) Logistic Regression Model without smote, (b) Precision-Recall Curve (AUC) Logistic Regression Model with smote and 30\% Minority, (c) Precision-Recall Curve (AUC) Logistic Regression Model with smote and 40\% Minority, (d) Precision-Recall Curve (AUC) Logistic Regression Model with smote and 50\% Minority.}
\label{qsmote_Fig20}
\end{figure}

In the next section we will describe the impact of SMOTE on the above evaluation charts.

\subsubsection{Impact of Quantum SMOTE on Model Statistics}
As we applied SMOTE on our two chosen models, we observed different behaviors of the models post-application of QuantumSMOTE.

\textbf{Random Forest:} \\
The Random Forest model excels in effectively addressing the Telecom Churn Dataset, particularly when dealing with imbalances in class distribution. The model's intrinsic advantages, together with its performance improvements using the SMOTE, provide a detailed analysis of its impact in tricky classification scenarios. As we walk through the model's performance parameters of Confusion matrices (Figs. \ref{qsmote_Fig13} and \ref{qsmote_Fig14}), Receiver Operating Characteristic Curve (AUC-ROC) (Fig. \ref{qsmote_Fig15}), Precision Recall Curve (AUC) (Fig. \ref{qsmote_Fig16}) we can see gradual improvements with induction of synthetic samples using SMOTE. We discuss the overall improvements in the points below.

\begin{itemize}
\item \textit{Performance Without Synthetic Data}

The introduction of SMOTE to the dataset led to observable improvements across various performance measures. Notably, as the percentage of synthetic minority increased, both test accuracy and F1 scores saw visible improvements. These improvements highlight the synergy between Random Forest's ensemble methodology and the balanced class distribution achieved through SMOTE. The model's adaptability to handle more balanced datasets and improve in predictive accuracy and precision recall underscores its versatility and effectiveness in handling imbalanced data scenarios.

\item Effects of Varying Degrees of Synthetic Data Augmentation on Performance

\textit{30\% Minority with Synthetic Data}: 

Test accuracy and F1 scores started to rise at this augmentation level, signaling the start of performance gains. With no change to the training data, the model achieved a test accuracy of 0.800813 and an F1 score improvement of 0.6343. Both the PR and ROC AUCs increased, reaching 0.757604 and 0.854414, respectively.

\textit{$40\%$ and $50\%$ Minority with Synthetic Data}: 

The test accuracy (0.822183) and F1 score (0.764202) were significantly improved by 40\% Minority using Synthetic Data. PR had an AUC of 0.888143 and ROC had an AUC of 0.905165.
The test accuracy increased to 0.846306 and the F1 score to 0.834755 with 50\% Minority using Synthetic Data. At their peak, PR and ROC AUC values were 0.940063 and 0.928649, respectively.
Both the 40\% and 50\% SMOTE augmentation levels improved the model more, but the 50\% augmentation level was when it really shone. Results showing significant improvements in test accuracy, F1 scores, and AUC scores for PR and ROC show that the model is better at identifying the minority class and can generalize more effectively.
\end{itemize}

\textbf{Logistic Regression:}
Performance in the analysis of the Logistic Regression model depicts its ability to handle class imbalance, especially when augmented with the SMOTE. We describe the behavior of Logistic Regression and its outcomes across different scenarios in following sections based on Confusion matrices (Figs. \ref{qsmote_Fig17} and \ref{qsmote_Fig18}), Receiver Operating Characteristic Curve (AUC-ROC) (Fig. \ref{qsmote_Fig19}), Precision-Recall Curve (AUC) (Fig. \ref{qsmote_Fig20}).

\begin{itemize}
\item \textit{Performance Without Synthetic Data:} Initially, the Logistic Regression model showed decent performance with a test accuracy of 0.796622, indicating its ability to accurately predict outcomes in over 80\% of cases.

Nevertheless, the F1 score, which is calculated as the harmonic mean of accuracy and recall, had a relatively low value of 0.523878. This suggests that while the model was usually reliable, it had challenges in achieving a trade-off between accuracy and recall, especially in correctly identifying the minority class. The Precision-Recall (PR) and Receiver Operating Characteristic(ROC) obtained Area Under the Curve (AUC) scores of 0.60415 and 0.814921, respectively. These scores indicate a reasonable potential to differentiate between classes, while there is potential for improvement in managing unbalanced data.

\item \textit{30\% Minority with Synthetic:} By inducing synthetic data to constitute 30\% of the minority class, the test accuracy saw a slight decline to 0.759485. This reduction implies that while the synthetic data was intended to balance the distribution of classes, it could have contributed to the complexity of class distribution that somewhat affected the general accuracy of predictions. However, the F1 score saw a small rise to 0.537158, suggesting that the model's capacity to maintain a balance between accuracy and recall improved under somewhat more equitable class settings. The AUC scores for PR (Precision-Recall) and ROC (Receiver Operating Characteristic) saw marginal enhancements to 0.632638 and 0.81238, respectively. These gains indicate a minor boost in the model's ability to differentiate between the classes when synthetic data is employed.

\item \textit{40\% Minority with Synthetic:} With the percentage of synthetic data was increased to 40\%,the test accuracy decreased to 0.700469. However, the F1 score increased to 0.607626. This implies that while the model's overall prediction accuracy declined, its capacity to detect the minority class improved, as shown by the higher F1 score. The area under the curve (AUC) scores for precision-recall (PR) and receiver operating characteristic (ROC) were 0.673914 and 0.769356, respectively. These values suggest that the model's accuracy and recall balance improved, but there was a minor decline in its overall discriminating power.

\item \textit{50\% Minority with Synthetic:} By using synthetic data to achieve a 50\% minority representation, the model demonstrated a notable improvement in test accuracy, reaching 0.733763. Yet, the F1 score increased substantially to 0.742446. The substantial rise in the F1 score demonstrates the improved ability of the model to properly detect the minority class due to a more evenly balanced dataset. The area under the curve (AUC) scores for precision-recall (PR) and receiver operating characteristic (ROC) increased to 0.778797 and 0.807275, respectively, indicating the enhanced ability of the model to distinguish between classes in a more balanced setup. 
\end{itemize}

\definecolor{Parsley}{rgb}{0.152,0.325,0.09}
\definecolor{TitanWhite}{rgb}{0.988,0.988,1}
\definecolor{MintTulip}{rgb}{0.752,0.901,0.96}
\definecolor{WePeep}{rgb}{0.949,0.807,0.937}
\definecolor{Concrete}{rgb}{0.949,0.949,0.949}
\definecolor{Skeptic}{rgb}{0.827,0.925,0.862}
\definecolor{BlizzardBlue}{rgb}{0.643,0.858,0.945}
\definecolor{LightOrchid}{rgb}{0.909,0.65,0.882}
\definecolor{Texas}{rgb}{0.964,0.964,0.6}
\definecolor{MossGreen}{rgb}{0.623,0.839,0.686}
\definecolor{TurquoiseBlue}{rgb}{0.4,0.756,0.905}
\definecolor{Orchid}{rgb}{0.866,0.494,0.827}
\definecolor{GoldenFizz}{rgb}{0.988,0.988,0.192}
\definecolor{Fern}{rgb}{0.388,0.745,0.482}
\definecolor{PictonBlue}{rgb}{0.266,0.701,0.882}
\definecolor{Orchid1}{rgb}{0.847,0.427,0.803}
\definecolor{RoyalPurple}{rgb}{0.439,0.188,0.627}
\definecolor{DeYork}{rgb}{0.45,0.772,0.537}
\definecolor{BlizzardBlue1}{rgb}{0.725,0.89,0.956}
\definecolor{WePeep1}{rgb}{0.933,0.749,0.917}
\definecolor{Yellow}{rgb}{0.996,0.996,0.054}
\definecolor{Cornflower}{rgb}{0.568,0.827,0.933}
\definecolor{LightOrchid1}{rgb}{0.909,0.658,0.886}
\definecolor{FringyFlower}{rgb}{0.686,0.866,0.741}
\definecolor{GoldenFizz1}{rgb}{0.988,0.988,0.16}
\begin{table*}
\centering
\resizebox{\linewidth}{!}{%
\begin{tblr}{
  width = \linewidth,
  colspec = {Q[377]Q[117]Q[117]Q[138]Q[88]Q[88]},
  row{1} = {Parsley,c,fg=white},
  row{2} = {c},
  row{8} = {RoyalPurple,c,fg=white},
  column{3} = {c},
  column{6} = {c},
  cell{1}{1} = {c=6}{0.924\linewidth},
  cell{2}{2} = {c=2}{0.234\linewidth},
  cell{2}{5} = {c=2}{0.176\linewidth},
  cell{3}{2} = {c},
  cell{3}{4} = {c},
  cell{3}{5} = {c},
  cell{4}{2} = {c},
  cell{4}{3} = {TitanWhite},
  cell{4}{4} = {MintTulip,c},
  cell{4}{5} = {WePeep,c},
  cell{4}{6} = {Concrete},
  cell{5}{2} = {c},
  cell{5}{3} = {Skeptic},
  cell{5}{4} = {BlizzardBlue,c},
  cell{5}{5} = {LightOrchid,c},
  cell{5}{6} = {Texas},
  cell{6}{2} = {c},
  cell{6}{3} = {MossGreen},
  cell{6}{4} = {TurquoiseBlue,c},
  cell{6}{5} = {Orchid,c},
  cell{6}{6} = {GoldenFizz},
  cell{7}{2} = {c},
  cell{7}{3} = {Fern},
  cell{7}{4} = {PictonBlue,c},
  cell{7}{5} = {Orchid1,c},
  cell{7}{6} = {yellow},
  cell{8}{1} = {c=6}{0.924\linewidth},
  cell{9}{2} = {c},
  cell{9}{3} = {Fern},
  cell{9}{4} = {MintTulip,c},
  cell{9}{5} = {WePeep,c},
  cell{9}{6} = {yellow},
  cell{10}{2} = {c},
  cell{10}{3} = {DeYork},
  cell{10}{4} = {BlizzardBlue1,c},
  cell{10}{5} = {WePeep1,c},
  cell{10}{6} = {Yellow},
  cell{11}{2} = {c},
  cell{11}{3} = {TitanWhite},
  cell{11}{4} = {Cornflower,c},
  cell{11}{5} = {LightOrchid1,c},
  cell{11}{6} = {Concrete},
  cell{12}{2} = {c},
  cell{12}{3} = {FringyFlower},
  cell{12}{4} = {PictonBlue,c},
  cell{12}{5} = {Orchid1,c},
  cell{12}{6} = {GoldenFizz1},
  vlines,
  hline{1-2,4-13} = {-}{},
  hline{3} = {1-3,5-6}{},
}
\textbf{Random Forest} &  &  &  &  & \\
\textbf{Scores} & \textbf{Accuracy Score} &  &  & \textbf{AUC Score} & \\
\textbf{Data Set Type} & \textbf{Train} & \textbf{Test} & \textbf{F1 Score} & \textbf{PR} & \textbf{ROC}\\
Without   Synthetic & 1.000 & 0.784 & 0.575 & 0.627 & 0.811\\
30\% Minority   with Synthetic & 1.000 & 0.801 & 0.634 & 0.758 & 0.854\\
40\% Minority   with Synthetic & 0.996 & 0.822 & 0.764 & 0.888 & 0.905\\
50\% Minority   with Synthetic & 0.996 & 0.846 & 0.835 & 0.940 & 0.929\\
\textbf{Logistic Regression} &  &  &  &  & \\
Without   Synthetic & 0.797 & 0.766 & 0.524 & 0.604 & 0.815\\
30\% Minority   with Synthetic & 0.753 & 0.759 & 0.537 & 0.633 & 0.812\\
40\% Minority   with Synthetic & 0.724 & 0.700 & 0.608 & 0.674 & 0.769\\
50\% Minority   with Synthetic & 0.732 & 0.734 & 0.742 & 0.779 & 0.807
\end{tblr}
}\caption{Table comparing Accuracy, F1 and AUC score of Random Forest Model for telecom churn dataset without SMOTE, and post SMOTE with minority\% as 30\%, 40\%, and 50\%.}
\label{ModelEvaluationsParmeters}
\end{table*}

\definecolor{MidnightBlue}{rgb}{0,0.125,0.376}
\definecolor{Cornflower}{rgb}{0.65,0.788,0.925}
\definecolor{FreshEggplant}{rgb}{0.501,0,0.501}
\definecolor{Malibu}{rgb}{0.38,0.796,0.952}
\definecolor{PastelGreen}{rgb}{0.556,0.85,0.45}
\definecolor{Tacao}{rgb}{0.945,0.662,0.513}
\definecolor{Turquoise}{rgb}{0.2,0.8,0.8}
\definecolor{PastelGreen1}{rgb}{0.513,0.886,0.556}
\definecolor{WePeep}{rgb}{0.949,0.807,0.937}
\definecolor{WineBerry}{rgb}{0.317,0.082,0.29}
\definecolor{LightOrchid}{rgb}{0.894,0.619,0.866}
\definecolor{GrannySmithApple}{rgb}{0.709,0.901,0.635}
\definecolor{Manhattan}{rgb}{0.968,0.78,0.674}
\definecolor{BrightTurquoise}{rgb}{0,1,0.8}

\begin{table}[tbp]
\centering
\resizebox{\linewidth}{!}{%
\begin{tblr}{
  width = \linewidth,
  colspec = {Q[248]Q[90]Q[90]Q[79]Q[79]Q[86]Q[69]Q[79]Q[79]},
  cells = {c},
  row{1} = {MidnightBlue,fg=white},
  row{3} = {PastelGreen},
  row{5} = {PastelGreen1},
  row{7} = {WineBerry},
  row{8} = {GrannySmithApple},
  row{10} = {GrannySmithApple},
  cell{1}{1} = {c=9}{0.898\linewidth},
  cell{2}{1} = {r=5}{Cornflower,fg=FreshEggplant},
  cell{2}{2} = {c=2}{0.18\linewidth},
  cell{2}{4} = {c=2}{0.158\linewidth},
  cell{2}{6} = {c=2}{0.155\linewidth},
  cell{2}{8} = {c=2}{0.158\linewidth},
  cell{3}{2} = {Malibu},
  cell{3}{3} = {Malibu},
  cell{3}{6} = {Tacao},
  cell{3}{7} = {Tacao},
  cell{3}{8} = {Turquoise},
  cell{3}{9} = {Turquoise},
  cell{5}{2} = {Malibu},
  cell{5}{3} = {Malibu},
  cell{5}{6} = {Tacao},
  cell{5}{7} = {Tacao},
  cell{5}{8} = {Turquoise},
  cell{5}{9} = {Turquoise},
  cell{7}{1} = {r=5}{WePeep,fg=blue},
  cell{7}{2} = {c=8}{0.651\linewidth},
  cell{8}{2} = {LightOrchid},
  cell{8}{3} = {LightOrchid},
  cell{8}{6} = {Manhattan},
  cell{8}{7} = {Manhattan},
  cell{8}{8} = {BrightTurquoise},
  cell{8}{9} = {BrightTurquoise},
  cell{10}{2} = {LightOrchid},
  cell{10}{3} = {LightOrchid},
  cell{10}{6} = {Manhattan},
  cell{10}{7} = {Manhattan},
  cell{10}{8} = {BrightTurquoise},
  cell{10}{9} = {BrightTurquoise},
  vlines,
  hline{1-2,7,12} = {-}{},
  hline{3-6,8-11} = {2-9}{},
}
\textbf{Confusion Matrix Comparison} &  &  &  &  &  &  &  & \\
\textbf{Random Forest} & \textbf{W/O Synthetic} &  & \textbf{30\% SMOTE} &  & \textbf{40\% SMOTE} &  & \textbf{50\% SMOTE} & \\
 & \textbf{TP} & \textbf{FP} & \textbf{TP} & \textbf{FP} & \textbf{TP} & \textbf{FP} & \textbf{TP} & \textbf{FP}\\
 & 899 & 114 & 931 & 96 & 1135 & 123 & 932 & 108\\
 & \textbf{FN} & \textbf{TN} & \textbf{FN} & \textbf{TN} & \textbf{FN} & \textbf{TN} & \textbf{FN} & \textbf{TN}\\
 & 202 & 192 & 189 & 260 & 236 & 636 & 192 & 785\\
\textbf{Logistic Regression} &  &  &  &  &  &  &  & \\
 & \textbf{TP} & \textbf{FP} & \textbf{TP} & \textbf{FP} & \textbf{TP} & \textbf{FP} & \textbf{TP} & \textbf{FP}\\
 & 807 & 116 & 915 & 112 & 998 & 260 & 706 & 334\\
 & \textbf{FN} & \textbf{TN} & \textbf{FN} & \textbf{TN} & \textbf{FN} & \textbf{TN} & \textbf{FN} & \textbf{TN}\\
 & 213 & 181 & 243 & 206 & 387 & 494 & 203 & 774
\end{tblr}
}\caption{Table comparing Confusion matrix of Random Forest and Logistic regression without SMOTE, and post SMOTE with minority\% as 30\%, 40\%, and 50\%.}
\label{ConfusionMatrixComparison}
\end{table}

\subsubsection{Final thoughts on SMOTE Performance}
The comparison of Logistic Regression and Random Forest models, enhanced with SMOTE, demonstrates the intricate nature of resolving class imbalance in machine learning. The performance enhancements of the Logistic Regression model, particularly in achieving a balanced precision-recall trade-off with the use of SMOTE, are consistent with the research conducted by Chawla et al. (2002). In their study, SMOTE was presented as a method to increase classifier performance by mitigating the issue of class imbalance via the generation of synthetic samples.

The Random Forest model demonstrates good performance, regardless of SMOTE. This underscores the model's intrinsic abilities in effectively dealing with class imbalances \cite{breiman_random_2001}. The ensemble strategy of the model, which combines predictions from numerous decision trees, inevitably offers a degree of resilience to imbalance, which is further strengthened by the use of SMOTE. Fernandez et al. \cite{fernandez_smote_2018} provide evidence supporting the effectiveness of ensemble approaches in handling unbalanced data. They propose that combining techniques such as Random Forest with SMOTE may lead to substantial improvements in model performance. All of the findings described in the assessment of Model performances are summarized in the table \ref{ModelEvaluationsParmeters} and the Confusion Matrix comparison table \ref{ConfusionMatrixComparison}.

\subsection{Inferences from Simulation} \label{Inferences}
In the process of creating the Quantum-SMOTE algorithm, we have come across several conclusions that we want to outline in the points below.

\begin{itemize}
\item The QuantumSMOTE algorithm functions similarly to the traditional SMOTE method but has the benefit of quantum phenomena.

\item The QuantumSMOTE technique utilizes the swap test and quantum rotation, distinguishing it from the standard SMOTE algorithm that relies on K Nearest Neighbors (KNN) \cite{Cover_KNN_1967, Altman_KNN_1992} and Euclidean distances \cite{chawla_smote_2002,blagus_smote_2013, han_borderline-smote_2005, Smote_enn_batista_2004}.

\item The QuantumSMOTE technique utilizes quantum rotation to eliminate neighbor dependencies and create several synthetic data points from a single data point in the minority class.

\item The technique includes hyperparameters that enable users to manage various elements of synthetic data creation, such as rotation angle, minority percentage, and splitting factor.

\item The QuantumSMOTE procedure generates synthetic data points to ensure that the distribution of variables closely resembles the original data distribution.

\item By selecting a smaller angle of rotation, the synthetic data points are positioned near the original minority data point, increasing the density of minority data points in a sparsely populated area.

\item The rotation circuit for minority data points does not encourage the use of any entanglement process or similar gates such as CNOT, ZZ, etc., since they will generate undesired effects on rotation and result in unexpected outcomes.

\item By using the compact swap test approach, more columns may be stored in fewer qubits. We used 5 qubits to handle 32 variables, and by scaling, we can handle 1024 variables with just 10 qubits.

\item The algorithm's use of low-depth circuits makes it less susceptible to issues associated with lengthy circuits like noise and decoherence. It effectively showcases how quantum techniques may enhance traditional machine-learning methods.

\item Similar to classical SMOTE, QuantumSMOTE generates synthetic data that enhances the Precision-Recall score of machine learning algorithms such as Logistic Regression \cite{Hosmer_2013_logistic} and significantly benefits ensemble algorithms like Random Forest \cite{breiman_random_2001}. This suggests its alignment with contemporary machine learning environments and confirms its applicability in current unbalanced classification scenarios.

\end{itemize}

\section{Conclusion}\label{Conclusion}

The QuantumSMOTE technique improves conventional class imbalance correction by employing quantum computing, particularly swap tests and quantum rotation, as opposed to classical approaches that rely on K Nearest Neighbors (KNN) and Euclidean distances. This quantum approach allows for the direct production of synthetic data points from minority class instances using quantum rotations, preventing the need for neighbor-based interpolation. QuantumSMOTE has customisable hyperparameters such as rotation angle, minority percentage, and splitting factor, allowing for personalised synthetic data synthesis to accurately solve dataset imbalances.

One notable feature of QuantumSMOTE is its capacity to generate synthetic instances that closely resemble the original data distribution, along with enhancing the balance of minority classes in datasets. The algorithm's use of compact swap tests enables efficient data representation, needing fewer qubits to manage a high number of variables, hence improving scalability and lowering quantum computing resource needs. Furthermore, its use of low-depth circuits reduces sensitivity to quantum noise and decoherence, making it a reliable option for quantum-enhanced data augmentation.

QuantumSMOTE's success is proven by its favorable influence on the Precision-Recall scores of machine learning algorithms such as Logistic Regression and Random Forest, highlighting its compatibility and utility in modern machine learning procedures. This technique is a forward-thinking integration of quantum computing with data science, providing an innovative and efficient solution to the problem of class imbalance in machine learning datasets.

\section*{Acknowledgment}
The authors are grateful to the IBM Quantum Experience platform and their team for developing the Qiskit platform and providing open access to their simulators for running quantum circuits and performing the experiments reported here. The authors also express gratitude towards the Center for Quantum Software and Information (CQSI) and Sydney Quantum Academy.

\section{Statements and Declarations}
\textbf{Competing Interests}: The authors have no financial or non-financial competing interests.\\
\textbf{Authors' contributions}:
The authors confirm their contribution to the paper as follows: 
Study conception and design: N.M., B.K.B., C.F., P.D.;\\ 
Data collection: N.M.;\\ 
Analysis and interpretation of results: N.M., B.K.B., C.F., P.D.;\\ 
Draft manuscript preparation: N.M., B.K.B., C.F., P.D.;\\
All authors reviewed the results and approved the final version of the manuscript.\\
\textbf{Funding}: Authors declare that there has been no external funding.\\
\textbf{Availability of data and materials}: All the data provided in this manuscript is generated during the simulation and can be provided upon reasonable request.

\bibliography{sn-bibliography}

\end{document}